\documentclass{aastex63}
\usepackage{hyperref}
\usepackage{makecell}
\usepackage{amsmath} 
\usepackage{cases}
\shorttitle{Resolved Debris Disk around HD 206893}
\shortauthors{Nederlander et al.}
\received{Jun 17, 2020}
\accepted{Jan 11, 2021}
\begin{document}

\title{Resolving Structure in the Debris Disk around HD 206893 with ALMA}

\author{Ava Nederlander}
\affiliation{Astronomy Department and Van Vleck Observatory, Wesleyan University, Middletown, CT 06459, USA}

\author{A. Meredith Hughes}
\affiliation{Astronomy Department and Van Vleck Observatory, Wesleyan University, Middletown, CT 06459, USA}

\author{Anna J. Fehr}
\affiliation{Astronomy Department and Van Vleck Observatory, Wesleyan University, Middletown, CT 06459, USA}

\author{Kevin M. Flaherty}
\affiliation{Department of Astronomy and Department of Physics, Williams College, Williamstown, MA 01267, USA}

\author{Kate Y. L. Su}
\affiliation{Steward Observatory, University of Arizona, 933 North Cherry Avenue, Tucson, AZ 85721, USA}

\author{Attila Mo\'{o}r}
\affiliation{Konkoly Observatory, Research Centre for Astronomy and Earth Sciences, Konkoly Thege Mikl\'{o}s \'{u}t 15-17, H-1121 Budapest, Hungary}
\affiliation{ELTE E\"otv\"os Lor\'and University, Institute of Physics, P\'azm\'any P\'eter s\'et\'any 1/A, 1117 Budapest, Hungary}

\author{Eugene Chiang}
\affiliation{Department of Astronomy, University of California at Berkeley, CA 94720-3411, USA}
\affiliation{Department of Earth and Planetary Science, University of California at Berkeley, CA 94720-4767, USA}

\author{Sean M. Andrews}
\affiliation{Center for Astrophysics $\vert$ Harvard \& Smithsonian, 60 Garden St., Cambridge, MA 02138, USA}

\author{David J. Wilner}
\affiliation{Center for Astrophysics $\vert$ Harvard \& Smithsonian, 60 Garden St., Cambridge, MA 02138, USA}

\author{Sebastian Marino}
\affiliation{Institute of Astronomy, University of Cambridge, Madingley Road, Cambridge CB3 0HA, UK}

\begin{abstract}
Debris disks are tenuous, dusty belts surrounding main sequence stars generated by collisions between planetesimals.  HD~206893 is one of only two stars known to host a directly imaged brown dwarf orbiting interior to its debris ring, in this case at a projected separation of 10.4\,au.  Here we resolve structure in the debris disk around HD~206893 at an angular resolution of 0\farcs6 (24\,au) and wavelength of 1.3\,mm with the Atacama Large Millimeter/submillimeter Array (ALMA).  We observe a broad disk extending from a radius of $<51$\,au to $194^{+13}_{-2}$\,au.  We model the disk with a continuous, gapped, and double power-law model of the surface density profile, and find strong evidence for a local minimum in the surface density distribution near a radius of 70\,au, consistent with a gap in the disk with an inner radius of $63^{+8}_{-16}$\,au and width $31^{+11}_{-7}$\,au.  Gapped structure has been observed in four other debris disks -- essentially every other radially resolved debris disk observed with sufficient angular resolution and sensitivity with ALMA -- and could be suggestive of the presence of an additional planetary-mass companion.
\end{abstract}
\section{Introduction}

Debris disks are a common outcome of the star and planet formation process, and serve as a signpost of mature planetary systems.  Bright debris disks are observed around some 25-30\% of main sequence stars \citep{trilling08,thureau14,montesinos16,sibthorpe2018}.  The true incidence is likely to be higher, since the sensitivity of current observations limits our ability to detect debris disks to those systems that are at least an order of magnitude more luminous than the disk generated by the Solar System's Kuiper Belt \citep[][and references therein]{matthews2014,hughes2018}.  While the dust in debris disks is certainly worthy of study in its own right, it also presents an opportunity to trace the properties of planetary systems.  

As direct imaging searches for exoplanets have matured, a small number of systems have been discovered in which both directly imaged companions and debris are present \citep[e.g.,][]{marois2008,lagrange2009,rameau13,mawet15,konopacky2016,meshkat2017}. These systems are extremely valuable dynamical laboratories.  While debris dust is subject to a number of different forces -- including radiation pressure, stellar winds, gas drag, and gravity -- the largest grains, imaged at millimeter wavelengths by facilities like the Atacama Large Millimeter/submillimeter Array (ALMA), are effectively impervious to all of the major forces except for gravity and collisions \citep[e.g.,][]{su05,strubbe06,wyatt08,wilner11,lohne12}.  

One useful dynamical concept is that of the ``chaotic zone," which is a term from 3-body dynamics that refers to the radial extent around a planet within which stable orbits for test particles do not exist \citep{wisdom1980,lecar2001}. The extent of the chaotic zone depends on the mass of the planet and its semimajor axis and eccentricity.  For a planet sculpting a debris belt, the separation between the planet's location and the edge of the debris belt will be equal to the extent of the planet's chaotic zone \citep[e.g.,][]{quillen2006,chiang2009,mustill2012,morrison2015}. Therefore, by locating the edge of the debris belt and measuring its relationship to the observed location of a directly imaged planet, we can place a dynamical constraint on the mass of the planet.  Such dynamical constraints on directly imaged companions are valuable because otherwise the masses of directly imaged companions are typically estimated using models of the luminosity evolution of planets and brown dwarfs, which are typically poorly calibrated and sensitive to the often highly uncertain age of the system \citep[e.g.,][]{chabrier00,baraffe03}.  Meaningful dynamical constraints on the mass of a directly imaged companion have already been shown to be possible with data from ALMA and the Submillimeter Array (SMA) in the case of the HR 8799 system \citep{wilner2018}.  

HD 206893, an F5V star located 40.8\,pc from Earth \citep{gaia2016,gaia2018}, is one of two known systems in which a brown dwarf orbits interior to a debris ring \citep{milli2017}.  The other is HR 2562 \citep{konopacky2016}. The debris disk around HD 206893 has been previously detected by \citet{williams2006}, who did not have sufficient angular resolution to measure the disk structure.  The disk outer radius has been marginally resolved by {\it Herschel} at a wavelength of 70\,$\mu$m, and analysis of the Spectral Energy Distribution (SED) suggests an inner radius of $\sim$50\,au \citep{moor2011}.  However, previous studies of debris-bearing systems have found that there is significant scatter in the relationship between spatially resolved disk size and the size estimated from the SED \citep[e.g.,][]{booth2013,pawellek2014,morales2016}, rendering spatially resolved observations critical for studying the interplay between the companion and the disk inner edge.  

The companion orbits the star at a projected separation of $10.4\pm 0.1$\,au, which is clearly interior to the projected debris ring radius \citep{milli2017}.  While it has so far been detected in two epochs of observation 10 months apart, confirming common proper motion with the star, the constraints on its orbital properties are not yet strong, and its eccentricity and alignment with the disk plane are still highly uncertain.  The initial mass estimates of the companion based on its {\it H}-band flux range from 24 to 73\,M$_\mathrm{Jup}$, depending in part on the poorly constrained age of the system \citep[They assumed 0.2-2\,Gyr, later refined to $250_{-200}^{+450}$\,Myr by][]{delorme2017}.  
Its spectral type lies among L5-L9 field dwarfs, although it is currently the reddest known object among young and dusty L dwarfs in the field; it is not yet clear whether its large color excess arises from a dusty atmosphere, or whether it is the result of disk reddening \citep{milli2017}. Analysis of additional SPHERE data find a best-fit mass in the range of 15-30\,M$_\mathrm{Jup}$, while noting that the data are compatible with everything from a 12\,M$_\mathrm{Jup}$ companion at an age of 50\,Myr to a 50\,M$_\mathrm{Jup}$ Hyades-age brown dwarf \citep{delorme2017}.   Follow-up observations combining direct imaging constraints with radial velocity data suggest the presence of an additional $\sim$15\,M$_\mathrm{Jup}$ companion at a separation of 1.4--2.6\,au from the star \citep{grandjean2019}.   

Current knowledge is consistent with a scenario in which the companion truncates the disk, but a measurement of the disk inner radius is necessary to confirm this scenario.  Here we observe the disk with ALMA (Section~\ref{sec:observations}), examine its radial intensity profile (Section~\ref{sec:results}) and measure the location of the inner radius of the debris belt  (Section~\ref{sec:analysis}).  We also report a  detection of a local minimum of the surface density near a radius of 70\,au in the outer debris belt.  If the minimum is due to a gap, this would be the fourth debris disk to exhibit gapped structure after HD 107146 \citep{ricci2015}, HD 92945 \citep{marino2019}, and HD 15115 \citep{macgregor2019} -- a feature that appears to be common in debris disks observed with sufficient sensitivity and angular resolution to resolve substructure.  The presence of gapped structure has several possible explanations, with the most compelling being an additional planetary-mass companion (Section~\ref{sec:discussion}).  We summarize the conclusions of our investigation in Section~\ref{sec:conclusions}.    

\section{Observations}
\label{sec:observations}

We observed the HD 206893 debris disk in six scheduling blocks between June and September 2018 (ALMA project 2018.1.00193.S, PI Hughes).  Three different antenna configurations with a baseline lengths ranging from 15 to 1246\,m are included in the data.  There are four spectral windows, each 1.875\,GHz wide to provide maximum sensitivity to dust continuum emission.  The three continuum windows were centered on frequencies of 228.5, 216.5, and 214.5\,GHz, with channel spacings of 15.6\,MHz (20.3\,km\,s$^{-1}$).  One spectral window was centered on the rest frequency of the CO(2-1) molecular line (230.53800\,GHz) with a channel spacing of 976 kHz (1.27\,km\,s$^{-1}$).  Table 1 lists the dates, times, number of antennas, baseline lengths, time on-source, the average precipitable water vapor, synthesized beam geometry, and rms noise values for each scheduling block.  The quasar J2148+0657 was used as both the flux and passband calibrator for all scheduling blocks.  J2131-1207 was used as the gain calibrator for the two June tracks, and J2146-1525 was used as the gain calibrator for all other tracks.  

Calibration, reduction, and imaging were carried out using standard tasks from the Common Astronomy Software Applications (\texttt{CASA}) package \citep{mcmullin2007}, version 5.1.1-5. The statistical weights for each visibility were recalculated using the variance of visibilities on nearby points in the uv plane, as described in \citet{flaherty2017}.  The ALMA technical handbook states that the expected absolute flux calibration uncertainty should be 5\% to 10\% for these Band 6 observations. 

\begin{table*}[ht]
\centering
\caption{ALMA Observations of HD 206893}
\label{table:observations}
\begin{center}
\resizebox{\textwidth}{!}{
\begin{tabular}{lcccccccc}
\hline
Date/Time (UT) & \# Antennas & Baseline lengths & On-source time & Average pwv & Beam Major Axis &
 Beam Minor Axis &  Beam PA & rms noise \\
 &  & (m) & (min) & (mm) & ('') & ('') & ($^\circ$) & ($\mu$Jy\,beam$^{-1}$)\\
\hline
Jun27/06:52 & 46 & 15-312 & 57.2 & 1.3 & 1.62  & 1.26  & -79.0 & 12.8\\
Jun27/07:49 & 46 & 15-312 & 57.0 & 1.1 & 1.62  & 1.25  & -68.8 & 11.7\\
Aug30/02:23 & 45 & 15-782 & 68.9 & 1.7 & 0.73  & 0.57  & 76.8 & 12.5\\
Aug30/03:34 & 45 & 15.1-782 & 68.9 & 1.6 & 0.75  & 0.58  & 80.7 & 12.3\\
Sep10/01:22 & 46 & 15-1213 & 71.8 & 2.1 & 0.56  & 0.43  & 53.5 & 13.0\\
Sep17/01:22 & 45 & 15-1246 & 69.0 & 0.6 & 0.49  & 0.34  & 54.7 & 10.0\\
{ Combined} & \nodata & 15-1246 & 392.8 & \nodata & 0.71  & 0.58  & 66.6 & 5.5\\
\hline
\end{tabular}
}
\end{center}
\end{table*}

\section{Results}
\label{sec:results}

\begin{figure}[ht!]
\centering
\includegraphics[angle=0,width=\textwidth]{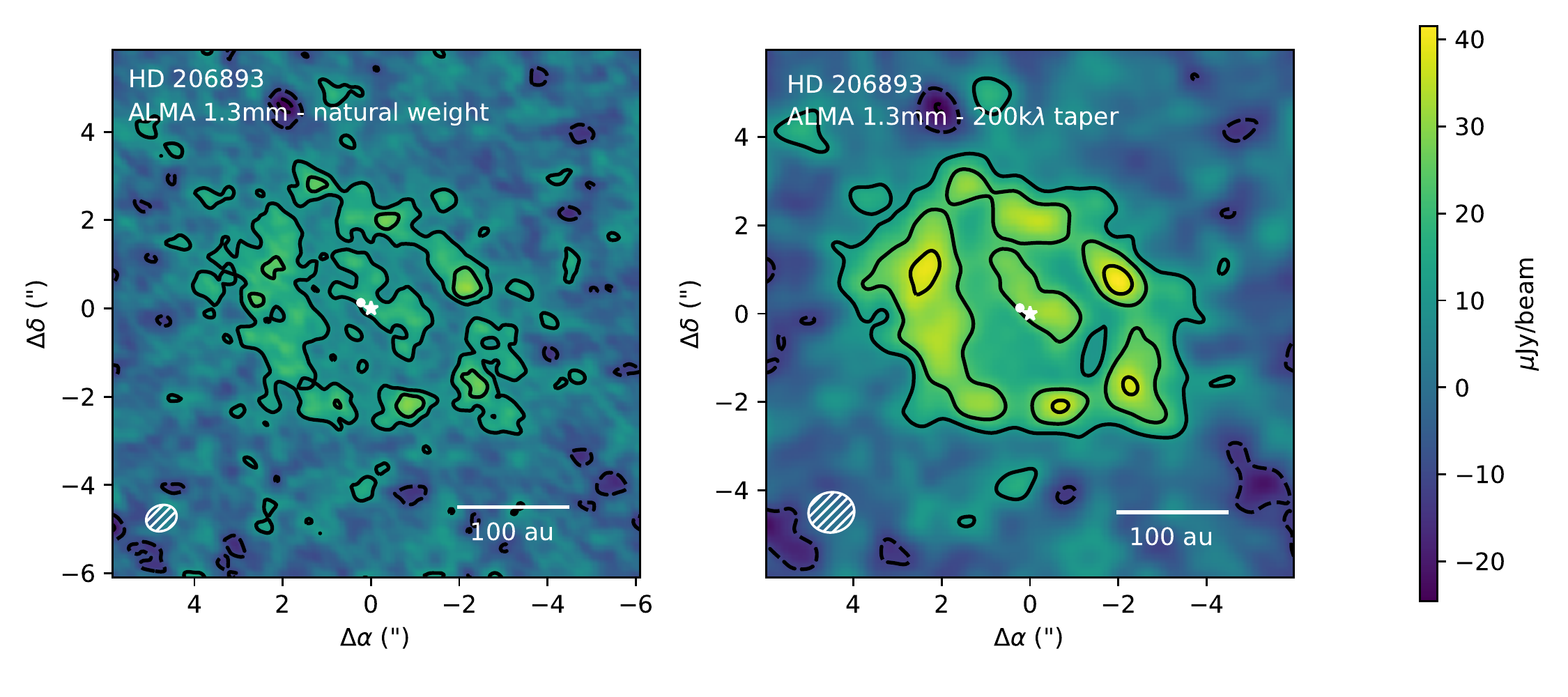}
\caption{(Left) Naturally weighted ALMA image of the 1.3\,mm continuum emission from the HD 206893 system.  (Right) Same image with a visibility-domain taper of 200\,k$\lambda$ applied to bring out the large-scale structure of the source.  In both panels, contour levels are [-2,2,4,6]$\times \sigma$, where $\sigma$ is the rms noise in the image: 5.5\,$\mu$Jy\,beam$^{-1}$ for the naturally weighted image and  6.0\,$\mu$Jy\,beam$^{-1}$ for the image with the taper. The hatched ellipse in the lower left corner represents the size and orientation of the synthesized beam: 0\farcs71$\times$0\farcs58 for the naturally weighted image and 0\farcs9$\times$1\farcs0 for the tapered image.  The star symbol represents the pointing center of the observations, i.e., the expected position of the star including a proper motion correction, and the dot represents the position of the brown dwarf companion directly imaged by \citet{milli2017}. }
\label{fig:images}
\end{figure}

The four spectral windows were combined to generate images of the dust continuum emission. Figure~\ref{fig:images} shows a naturally weighted image of the combined data set generated using the \texttt{CASA} task \texttt{tclean}.  We applied a 200\,k$\lambda$ taper to the interferometric data to bring out the large-scale structure of the debris ring. 

Using the \texttt{MIRIAD} task \texttt{cgcurs}, we measure an integrated flux density of $670\pm 30$\,$\mu$Jy at a wavelength of 1.3\,mm enclosed within the 3$\sigma$ contours of the tapered image {(not including the systematic flux calibration uncertainty)}.  Using the \texttt{CASA} \texttt{viewer} task, we measure the extent of the region enclosed by the 3$\sigma$ contours, which extends to a diameter of 7\farcs3 (corresponding to a radial extent of 150\,au) along the broadest dimension, and 5\farcs0 along the narrowest dimension.  The morphology of the disk traced by the best-detected (SNR$>$4) region can be described as an ellipse (i.e., an inclined ring) with some extra central flux in the interior.  There is a clear deficit of flux between the outermost bright ring and the central flux component, which exhibits a diameter along the major axis of approximately 2\farcs4 (corresponding to a radius of 49\,au).  

\begin{figure}[ht!]
\centering
\includegraphics[angle=0,scale=0.8]{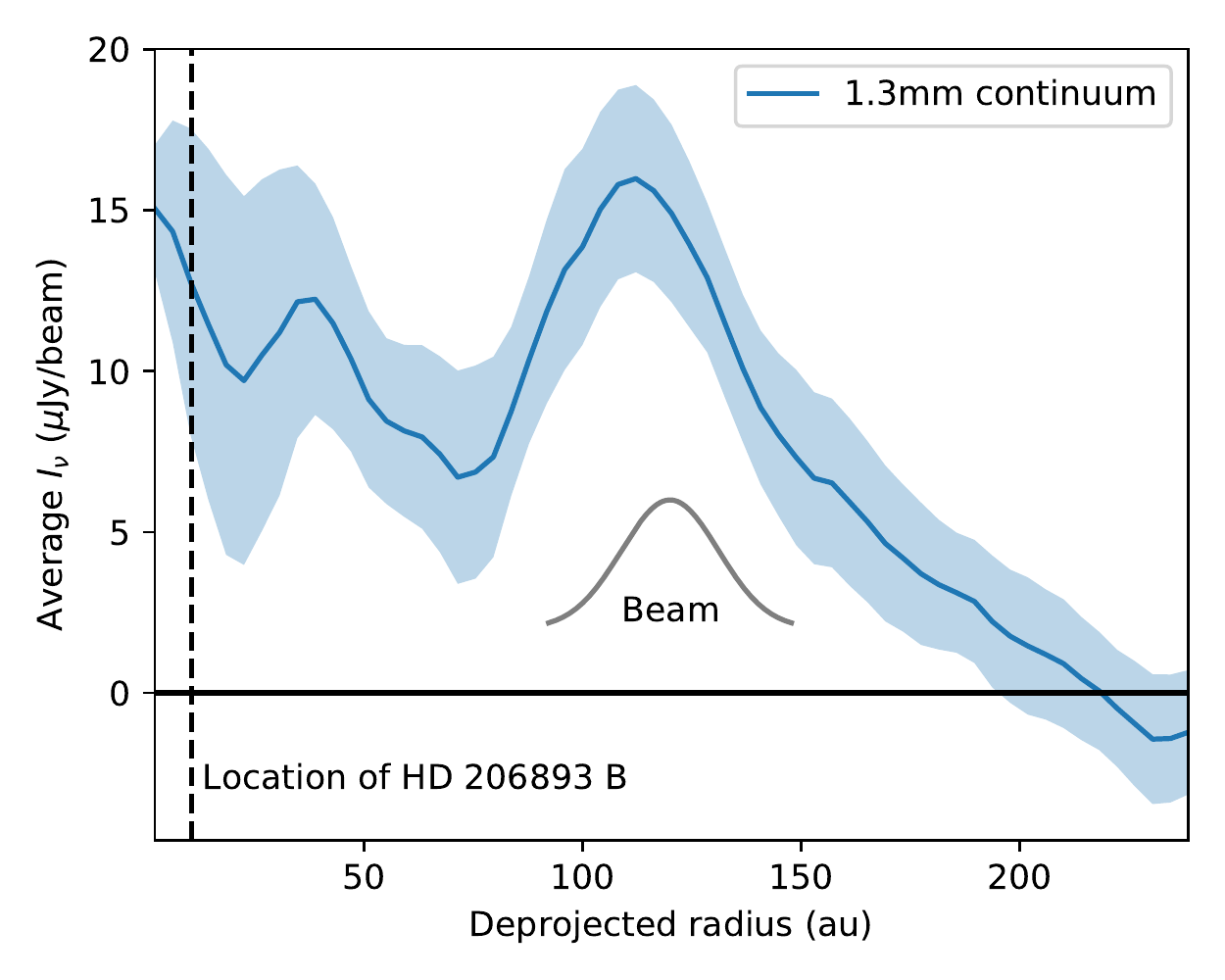}
\caption{Azimuthally averaged radial intensity profile {  of the naturally weighted image}, deprojected along elliptical contours assuming a circular geometry and an inclination to the line of sight of {  45$^\circ$ and position angle of 60$^\circ$}.  The blue shaded region represents the standard error of the mean for each radial bin, where the standard deviation is calculated for all pixels within the bin and then divided by the square root of the number of beams sampled.  The size of the synthesized beam is indicated by the gray Gaussian marked ``Beam."  The 10.4\,au projected separation of HD 206893 B \citep{milli2017} is marked with a vertical dashed line.  }
\label{figure:profile}
\end{figure}

Figure~\ref{figure:profile} plots the azimuthally averaged intensity profile as a function of linear separation from the central star, assuming that the disk has circular geometry and deprojecting the intensity profile along elliptical contours in the sky plane using the best-fit inclination of {45$^\circ$ and position angle of 60$^\circ$, which are the values obtained from the best-fit Markov Chain Monte Carlo (MCMC) model of a double power-law surface density profile with a radial gap, as described in Section~\ref{sec:analysis} below}.  The figure also shows the projected separation of the { brown dwarf} (BD) companion, HD 206893 B, as a dashed vertical line.  The intensity profile reveals a central peak that is broader than the synthesized beam (Full Width Half Maximum (FWHM) $\simeq26$\,au), indicating some disk flux located at radii close to the star.  There is a hint of an inner edge to the disk outside the orbit of the BD with a local peak at a radius of 38\,au, but the slight bump is small compared to the uncertainty on the flux density and is likely insignificant.  There is a local minimum in flux at a separation of 77\,au, and the brightest peak occurs at a separation of 115\,au -- the estimated {  relative} uncertainty on all of these radii is $\pm$5\,au, the average of the major axis and minor axis lengths of the synthesized beam divided by the average signal-to-noise ratio (SNR) of the disk of $\sim$5.  We further analyze the disk structure, including potential deviations from axisymmetry, in Section~\ref{sec:analysis} below.

The surface brightness of the central peak, $19.7\pm5.5$\,$\mu$Jy\,beam$^{-1}$, is approximately consistent with the expected flux density of 12.6\,$\mu$Jy for a star with a temperature of 6500\,K and radius 1.26\,R$_\odot$ \citep{delorme2017}, when approximating the star as a blackbody (in this case, the systematic flux uncertainty provides the largest source of error in the comparison).  {  While the properties of stars at millimeter wavelengths are not well understood, and can deviate significantly from the expected blackbody flux for a variety of reasons including chromospheric emission and flaring \citep{cranmer2013,white2018,white2019,white2020}, this estimate provides a ballpark figure to guide our interpretation of the emission morphology.}  As we demonstrate in Section~\ref{sec:analysis} below, there is a statistically significant contribution from a disk component located around the central point source that is only marginally resolved.  When the disk is modeled with a gap, which allows for an inner disk component that can contribute to the flux around the  point source, the stellar flux in the model is $17^{+5}_{-6}$\,$\mu$Jy, consistent with expectations -- but when the disk is modeled without a gap, the inner radius of the disk gets pushed to larger radii (presumably to avoid filling the gap with too much flux) and as a result the stellar flux in the model becomes elevated to $35^{+2}_{-3}$\,$\mu$Jy in an attempt to reproduce the flux from the inner regions of the debris belt.  The measured position of the central flux density peak is offset from the proper motion corrected {\it Gaia} position of the star by 0\farcs48, which is marginally significant at the 3$\sigma$ level (taking into account only the relative position uncertainty of the angular resolution $\theta$ divided by the SNR); however, subsequent analysis including detailed modeling of both the disk and the star with MCMC techniques (Section~\ref{sec:analysis} and Table~\ref{table:param_gap} below) yields no statistically significant difference between the expected and observed position of the star.  In this case the relative uncertainty is the larger source of uncertainty compared to the theoretical astrometric uncertainty of $\sim$40\,mas, calculated for the frequency, baseline length, and SNR of our observation based on the equation in Chapter 10.5.2 of the ALMA technical handbook. 

We did not detect any gas emission within a range of $\pm10$\,km\,s$^{-1}$ around the systemic velocity of the source, with a 3-sigma upper limit of approximately 40\,mJy\,km\,s$^{-1}$.  {  The upper limit was measured by generating a moment 0 map for the $\pm10$\,km\,s$^{-1}$ velocity range and then integrating the emission enclosed by the 2-sigma contours for the continuum emission imaged with the 200\,k$\lambda$ taper.}  We assumed a systemic velocity of -12.45\,km\,s$^{-1}$ in the heliocentric rest frame \citep{grandjean2019,gaia2018}, which translates to -5.07\,km\,s$^{-1}$ in the LSRK frame.  {  Our upper limit translates to a flux of $3.1\times10^{-22}$\,W\,m$^{-2}$, which is orders of magnitude above the predicted CO flux of $5.6\times10^{-25}$\,W\,m$^{-2}$ from \citet{kral2017}, indicating that the upper limit from this observation cannot place useful constraints on the composition of the molecular gas.}

\section{Analysis}
\label{sec:analysis}

In this section, we analyze the visibility data from the HD 206893 millimeter emission to characterize in detail the spatial distribution of dust in the system.  We adopt a modeling approach that combines a parametric ray-tracing code to generate synthetic model images of an axisymmetric disk with an MCMC fitting algorithm, allowing us to characterize the radial distribution of dust in the system. 

\subsection{Modeling Formalism}

To measure the disk structure, we generated synthetic model images of debris disks with varying geometries that were compared with the interferometric data. For each model image, we compared the data with the model in the visibility domain and calculated a $\chi^2$ metric to evaluate the goodness of fit between the model and the data.

To generate sky-projected model images of debris disks, we use a ray-tracing code described in \citet{flaherty2015}\footnote{https://github.com/kevin-flaherty/disk\_model3}, which is a {  Python} adaptation of an earlier IDL code by \citet{rosenfeld2013}.  The code translates parametric models of the density and temperature of dust and gas to a grid of density and temperature that is then rotated and integrated along the line of sight to produce a sky-projected map of the millimeter intensity.  While there is a fully 3-D version of the code available for eccentric disks, we use the 2-D version that assumes a circular, axisymmetric flux distribution with a linearly increasing scale height, since there is no {\it a priori} evidence for deviations from axisymmetry like a clear offset of the disk center from the star position (and indeed, the assumption is verified {\it a posteriori} by the lack of residuals after the best-fit model is subtracted from the data).  We assume a dust opacity of 2.3\,cm$^{2}$\,g$^{-1}$ \citep{beckwith1990}, yielding optically thin emission for the dust in all of the models within the parameter space explored by the MCMC chains.  Since density and temperature are degenerate for optically thin emission, we assume that these large grains are in blackbody equilibrium with the central star and calculate a dust grain temperature of 
\begin{equation}
    T_\mathrm{dust} = \left( \frac{L_*}{16 \pi \sigma R^2} \right)^{1/4}
\end{equation}
where $L_*$ is the bolometric luminosity of the star, $\sigma$ is the Stefan-Boltzmann constant, and $R$ is the distance of the dust grain from the central star.  We adopt a stellar luminosity of $L_*$ = 2.83\,L$_\sun$ for HD 206893 \citep{delorme2017}.  We {  assumed several different possible functional forms for the surface density of the disk; the functional forms for the surface density are summarized in Table~\ref{table:functional_forms}.} 

{  For example,} the surface density in the disk follows a power-law distribution where 
\begin{equation}
    \Sigma(R) = \Sigma_c R^p
\end{equation}
between radii of $R_\mathrm{in}$ and $R_\mathrm{out}$, where $\Sigma_c$ is a normalization for the total dust mass in the disk, $M_\mathrm{disk}$ (see Table~\ref{table:functional_forms} for definition), and $p$ is a power law index that we {  initially} set to a value of 1.0.  While the value of $p$ is interesting from the perspective of debris disk evolution, there is a well-known degeneracy between $p$ and the location of the outer radius $R_\mathrm{out}$ \citep[see, e.g., section 4.2.2. of][]{ricarte2013} and our data are of sufficiently limited sensitivity and angular resolution that we chose to focus on $R_\mathrm{out}$ rather than $p$.  We chose a value of 1 as a middle-of-the-road estimate bounded by theoretical predictions of 0 for collisional evolution models that assume a collisional lifetime of the largest planetesimal longer than the age of the system \citep{schuppler2016,marino2017}, and 7/3 for those that assume the opposite \citep{kennedy2010}.  

{  As it became clear that the local minimum and maximum evident in the radial surface density profile (Figure~\ref{figure:profile}) would require a more complex radial surface density distribution than the initial flat disk model, we explored two families of more complicated models: models with broken power laws that switched power law indices at one or two transition radii within the inner and outer radius of the disk, and models that included a sharp radial gap in the surface density profile at a particular radius.  When analyzing surface brightness features with modest signal-to-noise, it is often not possible to determine the exact shape of the feature, but broken power laws and sharp gaps are both common families of solutions assumed in the literature \citep[see, e.g.,][]{ricci2015} and adequately represent the extremes of an abrupt and deep gap vs. a broad and shallow gap.}  

After generating the sky-projected images, we {  apply a primary beam correction (multiplying by the primary beam)} and then convert the model image into synthetic model visibilities using the \texttt{MIRIAD} task \texttt{uvmodel} \citep{sault1995}, which we then compare with the data in the uv plane to calculate a $\chi^2$ value as a goodness-of-fit test.  Comparing data in the uv plane is desirable both because uncertainties are well characterized in the uv plane (unlike in the image domain, where the uncertainties are unknown and correlated between pixels, and nonlinearities can be introduced by the CLEAN process), and because it is agnostic to the choice of imaging parameters and allows us to take full advantage of the range of baseline lengths sampled.  

\begin{figure}[ht!]
\centering
\includegraphics[width=350pt, scale=0.30]{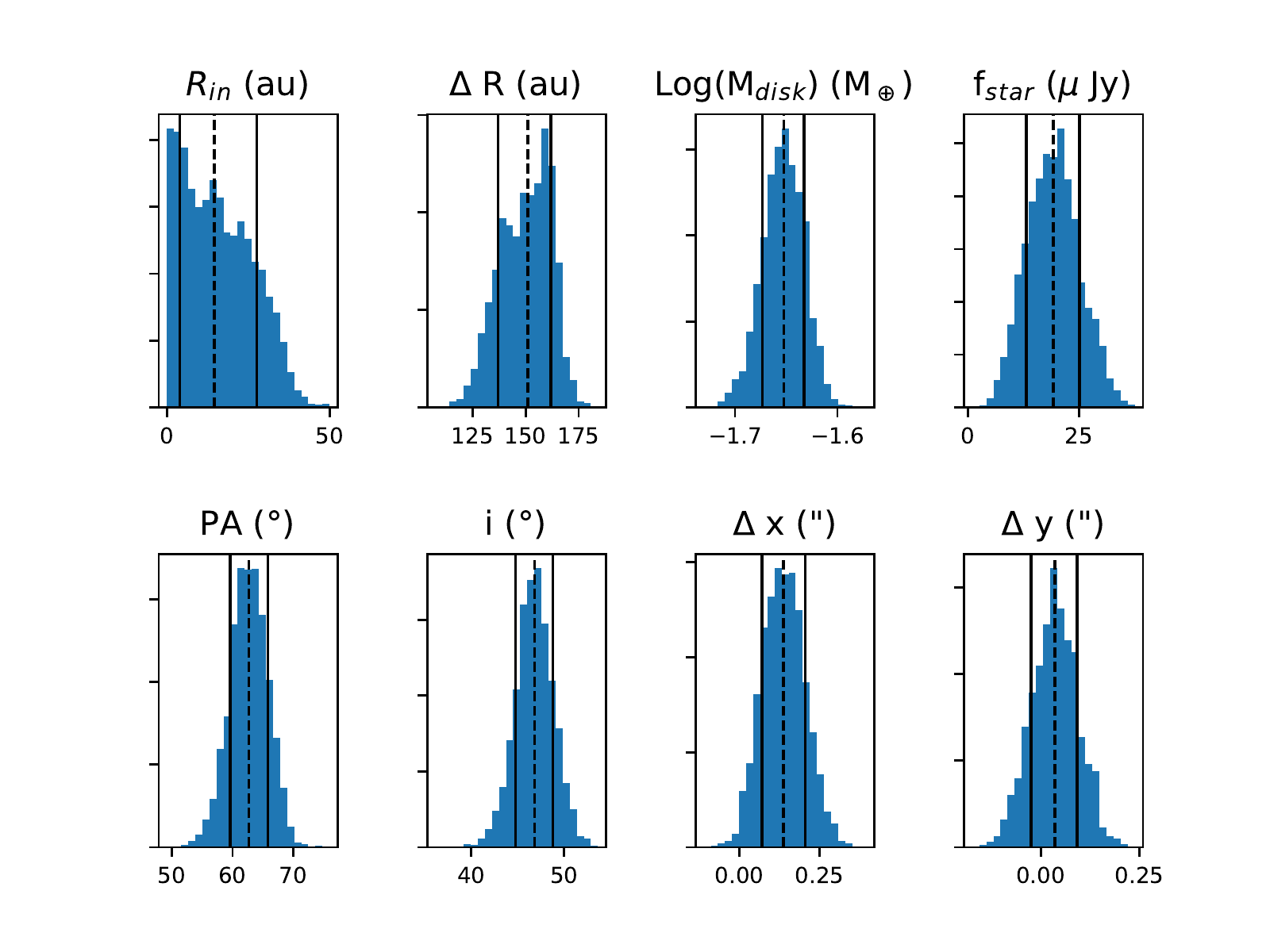}
\caption{{  Histograms of the marginalized posterior probability distributions for the flat disk MCMC run.  The central dashed line designates the median of each distribution while the outer lines mark the 16th and 84th percentiles.}}
\label{figure:histograms_continuous}
\end{figure}

\begin{figure}[ht!]
\centering
\includegraphics[angle=0,width=\textwidth, scale=0.30]{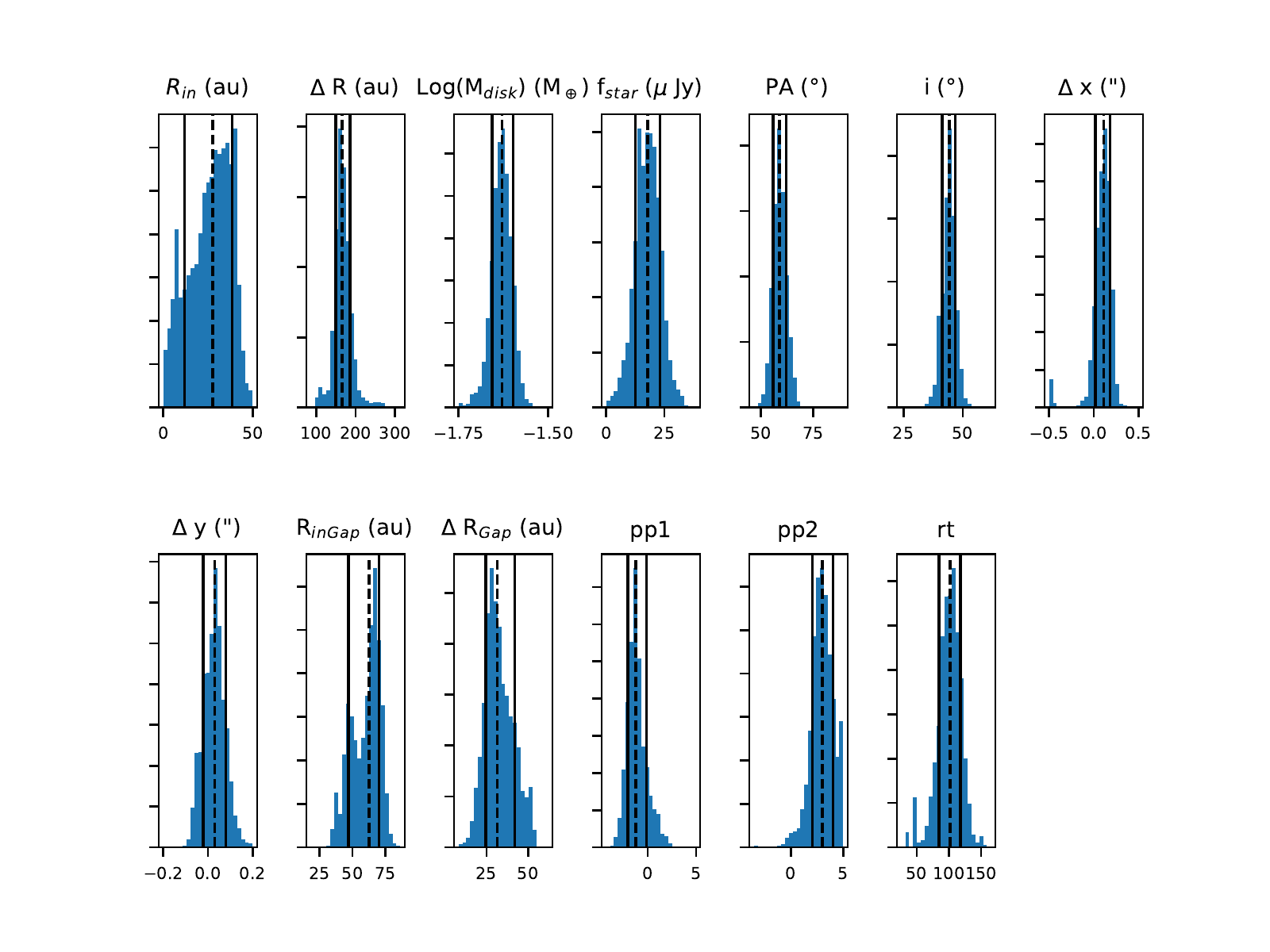}
\caption{Histograms of the marginalized posterior probability distributions for the {  double power law} MCMC run with a gap.  The central dashed line designates the median of each distribution while the outer lines mark the 16th and 84th percentiles.}
\label{figure:histograms_gap}
\end{figure}

\begin{figure}[ht!]
\centering
\includegraphics[angle=0,width=\textwidth, scale=0.30]{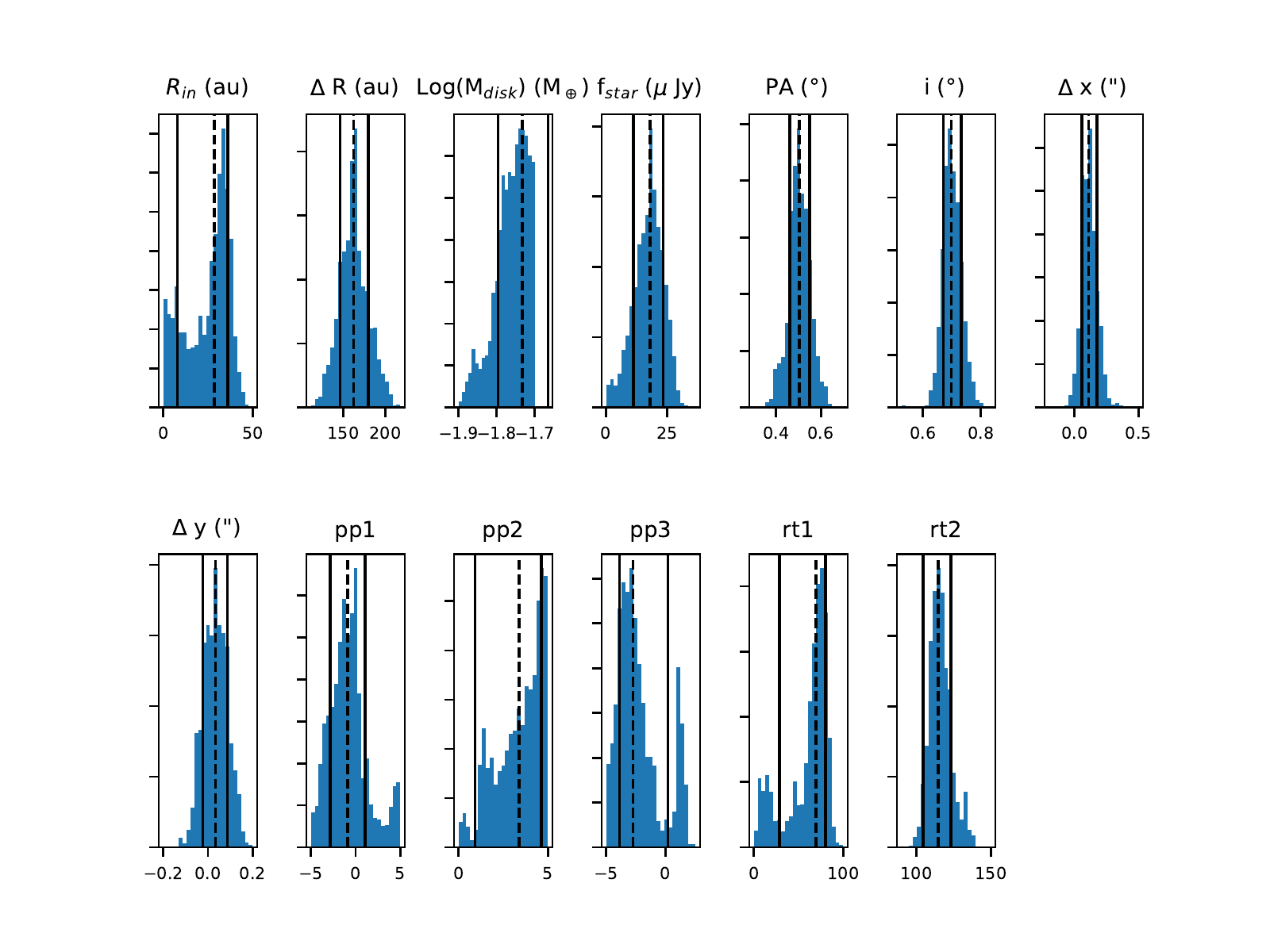}
\caption{Histograms of the marginalized posterior probability distributions for the {  triple power law} MCMC run without a gap.  The central dashed line designates the median of each distribution while the outer lines mark the 16th and 84th percentiles.}
\label{figure:histograms_nogap}
\end{figure}

We fit the models to the data using an affine-invariant MCMC sampler \citep{goodman2010} implemented in Python in the software package \texttt{emcee} \citep{foreman-mackey2013}. The goodness of a fit between the synthetic and observed visibilities are evaluated by a log-likelihood metric 
$\ln$ $\mathcal{L}$= $-\chi^2/2$. 
The MCMC code directs an ensemble of walkers in an exploration of parameter space, according to the calculated probability that a given walker position (representing a single set of model parameters) provides a better fit to the data than the previous walker position. After a ``burn-in" phase during which the walkers search downhill for the $\chi^2$ minimum, the process results in a set of model parameters describing the different ``steps" that each walker makes, which can then be agglomerated into a marginalized posterior probability distribution for each parameter (see Figs.~\ref{figure:histograms_continuous}, \ref{figure:histograms_nogap}, and \ref{figure:histograms_gap}).  

We performed several MCMC runs in order to investigate a variety of model formalisms.   Initially we varied eight parameters: the inner radius ($R_\mathrm{in}$), the distance between the inner part of the disk and the outer part of the disk ($\Delta$R, which is related to the disk outer radius as  $R_\mathrm{out} = R_\mathrm{in} + \Delta R$), the mass of the disk ($M_\mathrm{disk}$), the flux density of the central star $F_\mathrm{*}$, the position angle of the disk major axis (PA), the inclination of the disk relative to the observer's line of sight ($i$), and the position offset in right ascension ($\Delta x$) and declination ($\Delta y$) of the star-disk system relative to the pointing center of the interferometer. {  Subsequently, we introduced a gap inner radius ($R_\mathrm{in,Gap}$) and width ($\Delta R_\mathrm{Gap}$), and for the power law models, one or two transition radii ($Rt1$ and $Rt2$) with power law indices for the disk segments between the transition radii ($pp1$, $pp2$, and $pp3$).}
All parameters were sampled linearly except for disk mass which was sampled logarithmically, effectively equivalent to assuming a log-uniform prior, {  and position angle and inclination, which were sampled as $\cos{i}$ and $\cos{PA}$ to avoid undersampling of the extrema.} 

The initial run revealed two issues that made us suspect that a more complex distribution of material was necessary: first, the stellar flux was approximately double the anticipated value based on the blackbody approximation, and the models showed a much sharper central peak than we saw in the data, suggesting that there must be some diffuse emission around the central star.  In addition, the  models seemed to explore two regions of parameter space: initially they explored a region of parameter space with a small inner radius, which had a more reasonable stellar flux but left a ring of negative residuals farther from the star, and then eventually they landed in a region of parameter space where the disk inner radius was larger than expected -- too large for the inner edge to be carved by the brown dwarf, though it is certainly possible that other unseen companions could be carving the disk edge -- and the stellar flux was too high.  This behavior suggests that there might actually be two edges to the disk: one close to the star, accounting for the diffuse flux around the star, and another farther out.  Therefore, we explored a model with a radial gap in the disk.  We assumed uniform priors that required the inner radius of the gap to fall within the radial extent of the disk ([$R_{in}$,$R_{out}$]) and that required the width of the gap to be smaller than the total width of the disk ([0,$\Delta R$]).  The functional form of the gap is a top-hat: we assumed that the gap was completely empty, and the edges of the gap are step functions.  

{  In an attempt to remain agnostic about the functional form of the minimum in the surface brightness, we explored both a power-law with an empty gap, and a double power law, motivated by the set of functional forms assumed by \citet{ricci2015}.  However, we found that the double power law transition radius fell on the peak in the surface brightness around 115\,au, which meant that we were not able to evaluate whether a break in the power-law surface density could reproduce the observed surface brightness profile as well as a disk with a gap.  We therefore explored two additional profiles: a double power-law with a radial gap, and a triple power-law with two transitional radii.  All of the models were consistent in preferring a dip in the radial surface density profile near 75\,au and a peak near 115\,au, and the comparison between the latter two functional forms demonstrates that an empty gap with sharp edges yields comparable results to a more shallow power-law inflection point and is preferred with modest statistical significance.  Table~\ref{table:functional_forms} presents a summary of the functional forms, surface density normalization, free parameters, and best-fit lnprob values for each of the seven classes of models that we fit to the data.  For the remainder of the paper, we focus on the comparison between the flat disk (which ignores the local maximum and minimum of the surface density), the double power law with a gap, and the triple power law, since the latter two were the models that best (statistically and by eye) reproduced the features of the observations.} 

\begin{figure}[ht!]
\centering
\includegraphics[angle=0,width=\textwidth]{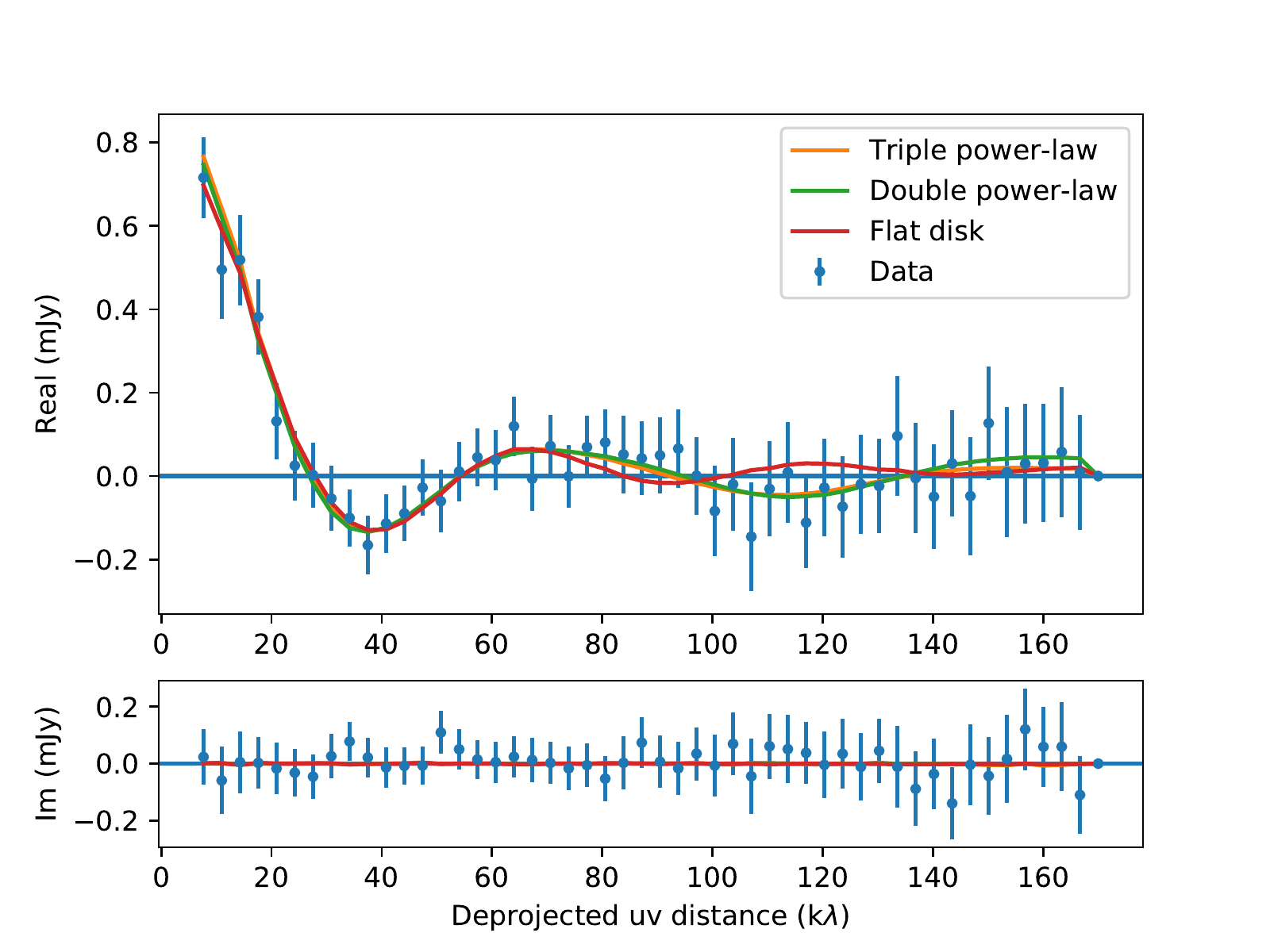}
\caption{Elliptically averaged visibility profile comparing the data (blue points) with the best-fit flat disk (red line), double power-law with a gap (green line) and triple power-law (orange line) models.  The top panel shows the real part of the visibilities while the bottom panel shows the imaginary part of the visibilities. The visibilities have been deprojected assuming an inclination of 44$^\circ$ and a position angle of 60$^\circ$, the best-fit values for the double power-law model.}
\label{figure:visibilities}
\end{figure}

The limits of the priors for all parameters are listed in Table~\ref{table:priors}.  {  For the flat disk we used 16 walkers and ran the chain for 2000 steps beyond the burn-in period.  For the double power law with gap we used 30 walkers and ran the chain for 2000 steps beyond the burn-in period.  For the triple power law, we used 30 walkers and ran the chain for 3000 steps beyond the burn-in period.  In all cases, the burn-in period was estimated by eye based on where the lnprob values leveled off to a relatively constant maximum.  We also followed up with an autocorrelation analysis showing that while the autocorrelation time was still rising by the end of each chain, all parameters had leveled off so that the fractional error in the mean was no more than a few percent, and the standard error of the mean was stable.  We show some sample plots from the double power-law model with a gap and the triple power-law model in the Appendix.}  The best-fit models, as well as the median and uncertainties given by the 16th and 84th percentiles of the posterior distribution, are presented in Table~\ref{table:MCMC}. Figure~\ref{figure:visibilities} shows the radially averaged visibility profile for the data, compared with the best-fit models for the flat disk, double power-law with gap, and triple power-law models.  The visibilities have been deprojected assuming an inclination of 44$^\circ$ and position angle of 60$^\circ$ (the best-fit values for the double power-law model).

\begin{deluxetable}{lccccc}
\centering
\label{table:priors}
\tablewidth{0pt}  
\tablecaption{MCMC Priors
}
\tablehead{
  \colhead{Parameters}  & \colhead{ Flat Disk} & \colhead{Double Power Law with Gap} &
 \colhead{Triple Power Law}}
\startdata
$R_{in}$\,(au) & [ 0, $1 \times 10^4$] & [0, $1 \times 10^4$] & [0, $1 \times 10^4$] \\
$\Delta$R(au) & [ 0, $1 \times 10^4$] & [0, $1 \times 10^4$] & [0, $1 \times 10^4$]\\
Log($M_\mathrm{disk}$) ($M_\earth$) & [ -10, -2] & [-10, -2] & [-10, -2]\\
$F_\mathrm{*}$\,($\mu$Jy) & [0, $1 \times 10^8$] & [0, $1 \times 10^8$] & [0, $1 \times 10^8$] \\
 cos(PA)\,($^\circ$) & [ -1, 1] & [ -1, 1] & [-1, 1]\\
cos($i$)\,($^\circ$) & [-1, 1] & [-1, 1] & [-1, 1]\\
$\Delta x$\,($\arcsec$) & [-5, 5] & [-5, 5] & [-5, 5]\\
$\Delta y$\,($\arcsec$) & [-5, 5] & [-5, 5] & [-5, 5]\\
$R_\mathrm{in,Gap}$\,(au) && [$R_\mathrm{in}$, $R_\mathrm{out}$] &\\
$\Delta R_\mathrm{Gap}$\,(au) && [0, $\Delta R$] &  \\
pp1\, && -5, 5 &  -5, 5 & \\
 pp2\, && -5, 5 &  -5, 5 & \\
 pp3\, && &  -5, 5 & \\
Rt1\, &&  $ R_{in}$, $R_{out}$ &  $R_{in}$, Rt2 & \\
Rt2\, &&&  Rt1, $R_{out}$ & \\
\enddata
\end{deluxetable}

\begin{figure}[ht!]
\centering
\includegraphics[angle=0,width=\textwidth]{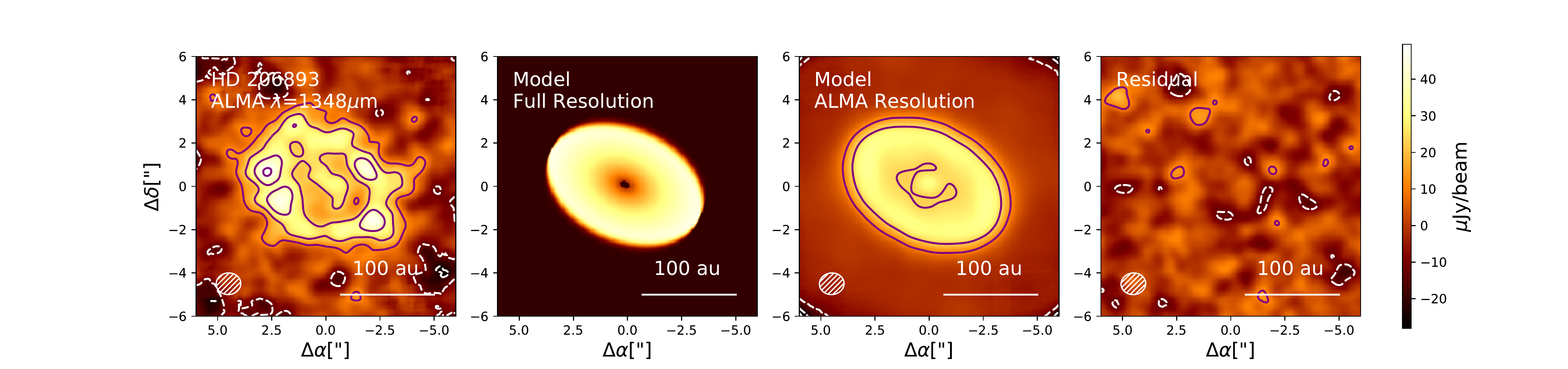}
\caption{{  (Left)} Naturally weighted ALMA image of the 1.3\,mm continuum emission from the HD 206893 system, with a taper of 200\,k$\lambda$ applied to bring out the large-scale structure of the source. {  (Center Left)} {  Full resolution model image for a flat disk showing the structure of the disk with a stellar flux equal to zero.} {  (Center Right)} Model image sampled at the same baseline lengths and orientations as the ALMA data, showing the best-fit model without a gap in the middle of the dust disk.  {  (Right)} Residual image after subtracting the model from the data in the visibility domain.  Contour levels and symbols are as in Figure~\ref{fig:images}. }
\label{figure:flat_disk}
\end{figure}

\begin{figure}[ht!]
\centering
\includegraphics[angle=0,width=\textwidth]{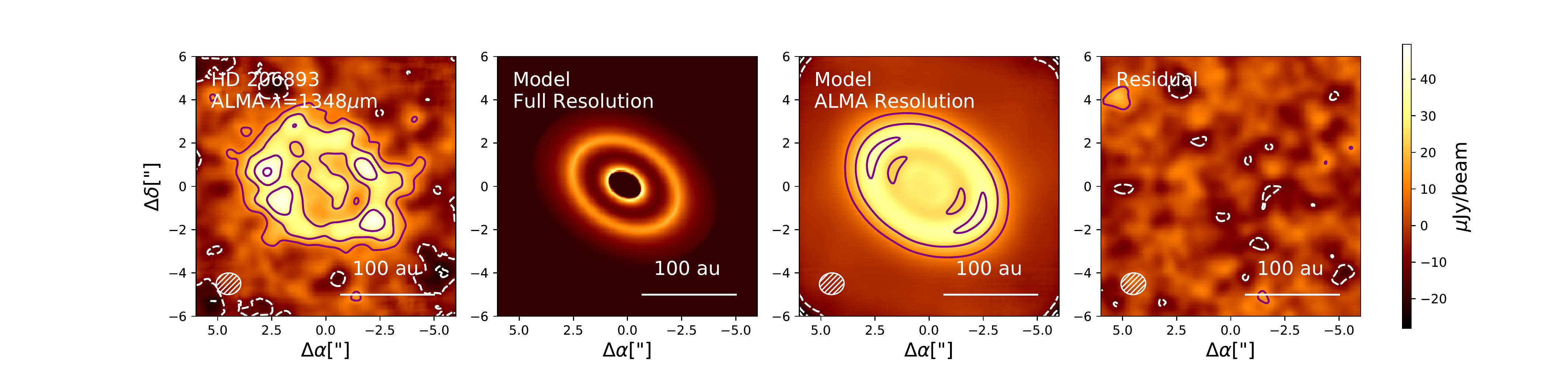}
\caption{{  (Left)} Naturally weighted ALMA image of the 1.3\,mm continuum emission from the HD 206893 system, with a taper of 200\,k$\lambda$ applied to bring out the large-scale structure of the source. {  (Center Left)} {  Full resolution image of a triple power-law model showing the structure of the disk with a stellar flux equal to zero.} {  (Center Right)} Model image sampled at the same baseline lengths and orientations as the ALMA data, showing the best-fit model with a gap in the middle of the dust disk.  {  (Right)} Residual image after subtracting the model from the data in the visibility domain.  Contour levels and symbols are as in Figure~\ref{fig:images}. }
\label{figure:no_gap}
\end{figure}

\begin{figure}[ht!]
\centering
\includegraphics[angle=0,width=\textwidth]{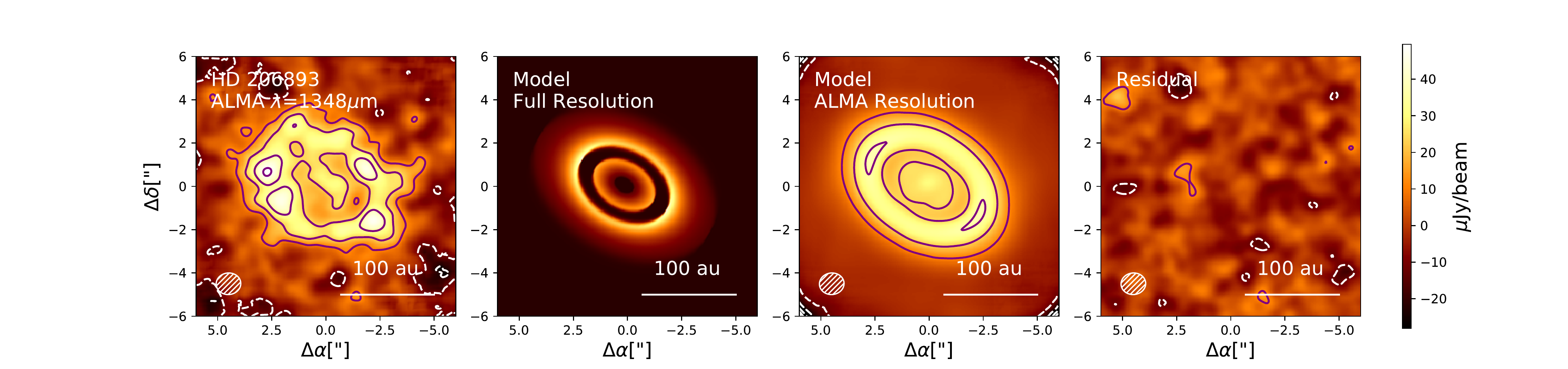}
\caption{Same as Fig~\ref{figure:no_gap}, but for a model of the disk that {  is parameterized as a double power law and} includes a gap in the radial dust distribution.}
\label{figure:with_gap}
\end{figure}

{  Figures~\ref{figure:flat_disk}, \ref{figure:no_gap}, and \ref{figure:with_gap}} show the tapered data image (left) compared with the best-fit model (center), sampled at the same baseline separations and orientations and imaged with the same parameters as the data, and the residuals (right).  The lack of significant ($>3\sigma$) residuals for these models shows that {  all models} adequately fit the data.  The main difference visible in the best-fit models is the distribution of flux around the star, towards the center of the disk.  For the model without a gap {  or power-law minimum}, the stellar flux is higher (by a factor of two) and the distribution of flux in the center of the system is strongly and centrally peaked. For the models {  with a gap or power-law break}, the stellar flux is lower (more in line with expectations from the blackbody estimate) and the emission is more diffuse around the star, since the model with a gap incorporates disk emission extending throughout the inner regions of the disk.

Since the models with a gap {  or power-law break} used {  five} more parameters than the model without a gap, we expect them to be able to provide a better fit simply due to the larger number of degrees of freedom.  In order to appropriately penalize the additional parameters when evaluating changes in the goodness-of-fit parameter, we used the AIC, a form of the Aikake Information Criterion, and the BIC, the Bayesian Information Criterion.  The BIC penalizes the use of additional parameters more than the AIC, and also takes into account the sample size (which the AIC does not).  A BIC score of $\Delta$BIC $>$ 10 implies ``very strong" evidence of a statistically improved fit \citep{kass1995}.  

{  When calculated, the $\Delta$BIC comparing the double power law with a gap to the flat disk is 37.6, implying ``very strong" evidence that the double power law with a gap is a better fit to the data than the flat disk.  The AIC also returns a probability of $1.1\times10^{-8}$ that the flat disk is a better fit than the model with a gap.  The corresponding BIC and AIC values for the triple power law gap compared with the flat disk model are 44.2 and $3.1\times10^{-7}$, also highly significant.  With so many models, there is a large number of potential comparisons we could make, but it is perhaps worth noting that for any given model type (single power law, or double power law) adding a radial gap provides a significantly better fit than the same model without the gap, even taking into account the increase in the number of parameters (the $\Delta$ BIC value for the flat power law with a gap compared to the flat power law without a gap is 16, the value for the single power law with a gap compared to the single power law without a gap is 23, and the value for the double power law with a gap compared to the double power law without a gap is 5.9). }  These values corroborate the conclusion that we have statistically significant evidence for a radial gap in the disk. 

{  As for the shape of the gap, the double power law with a gap and the triple power law both provide an excellent fit to the data using the same number of parameters, though the double power-law with a gap does have the larger lnprob value.  On the basis of the difference in lnprob, the probability that the double power law with a gap provides a better fit to the data than the triple power law is 0.03, which we take as suggestive but not conclusive evidence that a sharp, empty gap might provide a better fit to the data than a more shallow transition between two power laws. }


\begin{deluxetable}{lcccccc}
\centering
\label{table:MCMC}
\tablewidth{0pt}  
\tablecaption{MCMC Fitting Results \label{table:param_gap}
}
\tablehead{
    \hline
    \multicolumn{1}{l}{Parameter} & 
    \multicolumn{2}{c}{Flat Disk} &
    \multicolumn{2}{c}{Double Power Law with Gap} & 
    \multicolumn{2}{c}{Triple Power Law}\\
    \cline{2-3} \cline{4-5} \cline{6-7}
    &Best Fit & Median &
    Best Fit & Median &
    Best Fit & Median\\
 }
\startdata
$R_{in}$\,(au) & {  9} &  { $<44^a$} & { 21$^a$} & { $<51^a$} & { 35} & { $29^{+7}_{-20}$} \\
$\Delta$R(au) & {  155} & { $151^{+14}_{-11}$} & {  176} & { $166^{+20}_{-16}$} & { 159} & { $164^{+18}_{-14}$}\\
Log($M_\mathrm{disk}$) ($M_\earth$) & {  -1.66} & { $-1.63^{+0.03}_{-0.03}$} & {  -1.61} &  { $-1.63^{+0.03}_{-0.02}$} & { -1.74} & { $-1.74^{+0.07}_{-0.07}$} \\
$F_\mathrm{*}$\,($\mu$Jy) & {  16} & { $19^{+6}_{-6}$} & {  17} & { $18^{+5}_{-5}$} & { 14} & { $15^{+5}_{-6}$}\\
PA\,($^\circ$) & {  66} & { $63^{+3}_{-3}$} & { 60} & { $59^{+3}_{-3}$} & { 60} & { $63^{+3}_{-3}$}\\ 
$i$\,($^\circ$) & {  47} & { $47^{+2}_{-2}$} & { 45} & { $44^{+3}_{-3}$} & { 43} & { $47^{+3}_{-3}$}\\ 
$\Delta x$\,($\arcsec$) & {  0.11} & { $0.14^{+0.07}_{-0.07}$} & {  0.16} & { $0.11^{+0.07}_{-0.09}$} & { 0.11} & { $0.09^{+0.08}_{-0.09}$}\\
$\Delta y$\,($\arcsec$) & {  0.05} & { $0.04^{+0.06}_{-0.06}$} & {  0.04} & { $0.03^{+0.05}_{-0.05}$} & { 0.05} & { $0.06^{+0.11}_{-0.07}$}\\
$R_\mathrm{in,Gap}$\,(au) &&& {  67} & { $63^{+8}_{-16}$}& &\\
$\Delta R_\mathrm{Gap}$\,(au) &&& {  32} & { $31^{+11}_{-7}$}& &\\ 
{  pp1}\, &&& { -2.0} & { $-1.1^{+1.1}_{-0.8}$} & { -2.7} & { $-1.2^{+0.8}_{-1.1}$} \\
{  pp2}\, &&& { 2.8} & { $3.0^{+1.0}_{-0.9}$} & { 4.7} & { $>0.23$} \\
{  pp3}\, &&&  && { -3.7} & { $-3.0^{+1.3}_{-1.0}$} \\
{  Rt1}\, &&& { 97} & { $102^{+16}_{-17}$}  & { 73} & { $71^{+9}_{-33}$}  \\
{  Rt2}\, &&&  && { 113} & { $115^{+8}_{-7}$} \\
Ln prob & {  -10733015.1} & & {  -10732991.8} & & { -10732995.1} &\\
\enddata
\tablecomments{$^a$ The inner radius is unresolved in the models without a gap, so the best-fit value of {  9\,au or 21\,au} is not meaningful.  The upper limit of {  44\,au or 51\,au} represents the 99.7th percentile of the posterior distribution.}
\end{deluxetable}

\clearpage 


\begin{deluxetable}{lcccc}
\tablecaption{Summary of different functional forms assumed for the surface density profile
}
\label{table:functional_forms}
\tablehead{
  \colhead{Model Type} & \colhead{Surface Density Profile} & \colhead{Normalization} & \colhead{Variable Parameters} & \colhead{best-fit lnprob}}
\rotate
\startdata
\makecell{flat disk} & $\Sigma(r) = \Sigma_d r$  & 
$\Sigma_d = \frac{3 M_\mathrm{dust}}{2\pi (R_\mathrm{out}^{3} - R_\mathrm{in}^{3})}$
& \makecell{$R_{in}$, $\Delta$R, Log($M_\mathrm{disk}$), \\ $F_\mathrm{*}$, PA, $i$, $\Delta x$, $\Delta y$} & -10733015.1\\
\makecell{flat disk \\ with gap} & $\Sigma(r) = \Sigma_d r$ & $\Sigma_d = \frac{3 M_\mathrm{dust}}{2\pi (R_\mathrm{out}^{3} - R_\mathrm{in}^{3})}$ & \makecell{$R_{in}$, $\Delta$R, Log($M_\mathrm{disk}$), \\ $F_\mathrm{*}$, PA, $i$, $\Delta x$, $\Delta y$, \\ $R_\mathrm{in,Gap}$, $\Delta R_\mathrm{Gap}$} & -10733006.4\\
\makecell{power law} & $\Sigma(r) = \Sigma_d r^{pp}$ & \makecell{
$\Sigma_d = \frac{M_\mathrm{dust}(pp + 2)}{2\pi (R_\mathrm{out}^{2+pp} - R_\mathrm{in}^{2+pp})}$}
& \makecell{$R_{in}$, $\Delta$R, Log($M_\mathrm{disk}$), \\ $F_\mathrm{*}$, PA, $i$, $\Delta x$, $\Delta y$, \\ pp, rt} & -10733011.5 \\
\makecell{power law \\ with gap} & $\Sigma(r) = \Sigma_d r^{pp}$ & $\Sigma_d = \frac{M_\mathrm{dust}(pp + 2)} {2\pi (R_\mathrm{out}^{2+pp} - R_\mathrm{in}^{2+pp})}$
& \makecell{$R_{in}$, $\Delta$R, Log($M_\mathrm{disk}$), \\ $F_\mathrm{*}$, PA, $i$, $\Delta x$, $\Delta y$, $R_\mathrm{in,Gap}$, \\ $\Delta R_\mathrm{Gap}$, pp, rt} & -10733006.3\\
\makecell{double power law} & 
$ \Sigma(r) = 
\begin{cases}
\Sigma_t \times r^{-pp1} ~~ \mbox{if} ~~ R_\mathrm{in} < r < R_\mathrm{t}\\
\Sigma_t \times r^{-pp2} ~~ \mbox{if} ~~ R_\mathrm{t} < r < R_\mathrm{out}
\end{cases}
$
& $\Sigma_d = \frac{(M_\mathrm{dust}  j_1  j_2))} {2\pi (m_1 j_2 R_\mathrm{t}^{pp1} + m_2 j_1 R_\mathrm{t}^{pp2)}}$ & \makecell{$R_{in}$, $\Delta$R, Log($M_\mathrm{disk}$), \\ $F_\mathrm{*}$, PA, $i$, $\Delta x$, $\Delta y$, \\ pp1, pp2, rt} & -10733005.7\\
\makecell{double power law \\ with gap} & 
$ \Sigma(r) = 
\begin{cases}
\Sigma_t \times r^{-pp1} ~~ \mbox{if} ~~ R_\mathrm{in} < r < R_\mathrm{t}\\
\Sigma_t \times r^{-pp2} ~~ \mbox{if} ~~ R_\mathrm{t} < r < R_\mathrm{out}
\end{cases}
$
& $\Sigma_d = \frac{(M_\mathrm{dust}  j_1  j_2))} {2\pi (m_1 j_2 R_\mathrm{t}^{pp1} + m_2 j_1 R_\mathrm{t}^{pp2}}$  & \makecell{$R_{in}$, $\Delta$R, Log($M_\mathrm{disk}$), \\ $F_\mathrm{*}$, PA, $i$, $\Delta x$, $\Delta y$, $R_\mathrm{in,Gap}$, \\ $\Delta R_\mathrm{Gap}$, pp1, pp2, rt}  & -10732991.8\\
\makecell{triple power law}  & 
$ \Sigma(r) = 
\begin{cases}
\Sigma_{t1} \times r^{pp1} ~~ \mbox{if} ~~ R_\mathrm{in} < r < R_\mathrm{t1}\\
\Sigma_{t1} \times r^{pp2} ~~ \mbox{if} ~~ R_\mathrm{t1} < r < R_\mathrm{t2} \\
\Sigma_{t2} \times r^{pp2} ~~ \mbox{if} ~~ R_\mathrm{t2} < r < R_\mathrm{t3}
\end{cases}
$
& 
\makecell{$\Sigma_{t1} = \frac{M_\mathrm{dust} j_1 j_2 j_3}{2\pi (R_\mathrm{t1}^{-pp1} j_2 j_3 m_1 + R_\mathrm{t2}^{-pp2} j_1 j_3 m_2 + \frac{R_\mathrm{t2}^{pp2-pp3}}{R_\mathrm{t2}^pp2}j_1 j_2 m_3}$ \\
$\Sigma_{t2} = \Sigma_{t1} \left( \frac{R_{t2}}{R_{t1}}\right)^{pp2}$
}
& \makecell{$R_{in}$, $\Delta$R, Log($M_\mathrm{disk}$), \\ $F_\mathrm{*}$, PA, $i$, $\Delta x$, $\Delta y$,  \\  pp1, pp2, pp3, rt1, rt2} & -10732995.1\\
\enddata
\tablecomments{$j_1 = 2+pp1$, $j_2 = 2+pp2$, $j_3 = 2+pp3$, $m_1 = R_\mathrm{t1}^{j1} - R_\mathrm{in}^{j1}$, $m_2 = R_\mathrm{t2}^{j2} - R_\mathrm{t1}^{j2}$, $m_3 = R_\mathrm{out}^{j3} - R_\mathrm{t2}^{j3}$}
\end{deluxetable}

\clearpage

\section{Discussion}
\label{sec:discussion}

\subsection{Disk Structure Constraints}

The results of our MCMC fits demonstrate that the HD~206893 disk is well described by a broad underlying flux distribution extending from radii of {  $<51$\,au to $194^{+13}_{-2}$\,au} (this constraint on $R_\mathrm{out}$ is measured from the posterior distribution of $R_\mathrm{out} = R_\mathrm{in} + \Delta R$, which is not shown in Figure~\ref{figure:histograms_gap}), with a gap beginning at a radius of {  $63^{+8}_{-16}$\,au and exhibiting a width of $31^{+11}_{-7}$\,au}.  There are no statistically significant ($>3\sigma$) residuals, which indicates that the observations are well described by an azimuthally symmetric and circular distribution of flux.  Given the limited SNR ratio of the data (the brightest parts of the dust disk are approximately at the $5\sigma$ level on average), we can only rule out {  local (or point-to-point)} spatial variations in flux of $\gtrsim50$\% on spatial scales comparable to the beam FWHM of $\sim$30\,au or larger.  

{  When we compare our derived disk size to the sample of spatially resolved disks investigated by \citet{matra2018}, we note that the value of R for HD 206893 according to the definition laid out in that paper would be approximately 115 au.  Of their sample of ten F stars, only one (HR 8799) has a larger radius -- and that star has a particularly unusual $\lambda$ Boo spectrum that is in some ways more consistent with an A star spectral type.  The disk around HD 206893 is therefore among the largest known debris disks around F stars, although comparable to that around HD~170773 \citep{sepulveda2019}. }

Our model makes several simplifying assumptions that may or may not be realistic: the completely empty gap (rather than allowing the depth of the gap to vary), and the sharp edges of both the disk inner and outer radii and the inner and outer edges of the gap.  Better characterizing these features -- the depth of the gap, the sharpness of the edges -- would be interesting from the perspective of understanding debris disk evolution and interactions between the disk and the planet.  Unfortunately, this initial reconnaissance of the disk structure is necessarily limited in angular resolution and sensitivity, and the quality of the data do not justify a more complex model (as evidenced by the lack of residuals).  However, armed with this new knowledge of the critical spatial and flux scales for the system, future observations at higher angular resolution and sensitivity will be able to {  elucidate} these details.  

One of the key parameters of interest is the inner radius of the disk.  Previous models of the spatially unresolved SED had estimated that the location of the inner radius should fall around 50\,au \citep{moor2011}, estimated from the derived dust temperature of $49\pm2$\,K, although comparisons between spatially resolved observations and SED estimates tend to reveal a large amount of scatter, with spatially resolved measurements biased towards larger radii than SED-based estimates \citep{booth2013,pawellek2014,morales2016}.  In this case, the measured radius is smaller than the SED-estimated radius, and appears to be spatially unresolved with a radius of {  $<51$\,au}. This is good news for obtaining dynamical measurements of the mass of the brown dwarf: an inner radius of 50\,au \citep[as suggested by the SED models in][]{moor2011} is too large to be plausibly truncated by a brown dwarf with a projected separation of 10\,au, at least in the absence of extreme eccentricity, which is ruled out by the combination of astrometric, radial velocity, and direct imaging constraints explored by \citet{grandjean2019}. We should note that there is some tension between a derived inner radius of $<51$\,au and the 50\,K temperature constraint from the SED \citep{moor2011}, since even blackbody-like grains should be hotter than 50\,K at distances interior to $\sim$40\,au from the central star.  The temperature probed by the SED is likely dominated by the bulk of the opacity, i.e., $\sim \mu$m-sized grains in a collision-dominated debris disk. Small $\mu$m-sized grains are expected to locate further from the star than those of blackbody-like grains at the same temperature. The measured disk extent and the disk SED are largely consistent.  However, the morphology of the ALMA data seems to suggest that at least some millimeter-size grains must be present within the central 1-2 beams, corresponding to $\lesssim37$\,au in radius (assuming a beam FWHM of 0\farcs6 and a diameter of 1.5 beams for the central flux component).  Future investigation is needed, particularly the detailed shape of the infrared SED and the exact location of the inner disk edge.  

The bad news is that the current data do not have sufficient angular resolution to allow us to provide constraints on the mass of the BD companion that are more restrictive than the constraints from  \citet{grandjean2019}.  Our {  best-fit value for the inner radius of the triple power law model (35\,au)}, when interpreted in the context of chaotic zone theory and assuming a semimajor axis of 13\,au \citep[from the MCMC fits to the combined astrometric, RV, and direct imaging constraints in][]{grandjean2019}, provides an upper limit on the mass of the brown dwarf of $<1170$\,M$_\mathrm{Jup}$, assuming zero eccentricity.  The extent of the chaotic zone depends on the semimajor axis of the companion's orbit, as well as the mass ratio $\mu$ between the companion and the central star.  To derive the upper limit on the mass, we use the functional form derived by \citet{morrison2015} for the outer extent of the chaotic zone for a high-$\mu$ companion: $ \Delta a \simeq 1.7\mu^{0.31}a_p$.  It is possible to place somewhat better constraints through a more detailed dynamical analysis, although even that is difficult due to the likely presence of an additional companion causing the RV drift detected by \citet{grandjean2019}.  {  \citet{marino2020}} conduct just such an analysis and show that the location of the inner edge is consistent with being carved by the brown dwarf, though it could be farther out than expected based on chaotic zone theory due to secular resonances with the companion responsible for the RV drift.  Since the location of the inner edge of the disk is so steeply dependent on the companion mass, even a modest improvement in the angular resolution will yield substantially improved constraints on the dynamical mass measurement of the brown dwarf companion.  

Another important constraint provided by the circumstellar disk measurement is the inclination of the planetesimal belt.  While the nearly face-on configuration precludes meaningful constraints on the vertical structure, and therefore the dispersion of inclinations, the average inclination is constrained by our analysis to be {  $45^\circ \pm 4^\circ$}.  This range of inclinations is somewhat inconsistent with the plausible range of 20$^\circ$-41$^\circ$ for the brown dwarf companion reported by \citet{grandjean2019}, as well as the $30^\circ \pm5^\circ$ inclination of the stellar pole derived by \citet{delorme2017}. Interestingly, the version of the \citet{grandjean2019} dynamical MCMC analysis that combines the constraints from radial velocity, astrometry, stellar proper motion, and direct imaging results in a posterior distribution for the inclination of the companion of $45^\circ \pm 3^\circ$ and a mass of $10^{+5}_{-4}$\,M$_\mathrm{Jup}$, though they note that the median $\chi^2$ on the RV data is six times higher than when they do not include the stellar proper motion constraints, indicating that there may be some inconsistency in the ability of their model to reproduce both the RV and stellar proper motion data -- perhaps providing additional support for the presence of another companion in the system.  At this point, while we cannot rule out mutual inclination for the star and brown dwarf companion, coplanarity seems plausible.  

\subsection{Gap Detection and Implications}

The AIC and BIC comparison between the best-fit models provide strong evidence for a gap or local minimum in the surface density near a radius of 80\,au.  This marks the fourth detection of a radial gap in a broad debris belt, after HD 107146 \citep{ricci2015,marino2018}, HD 15115 \citep{macgregor2019}, and HD 92945 \citep{marino2019}.  While the sample size is not large, there are only a couple of other debris disks with broad ($\Delta R / R \gtrsim 1$) debris belts that have been imaged with sufficient resolution and sensitivity at millimeter wavelengths to detect substructure, including $\beta$ Pic \citep{dent2014,matra2019} and AU~Mic \citep{macgregor2013}, the latter of which in fact does exhibit tentative (AIC $3.7\sigma$, $\Delta$BIC = 4.3) evidence for a gap in its edge-on planetesimal belt \citep{daley2019}.  The newly detected gap in the HD 206893 debris disk is therefore part of an emerging trend of gapped structure in broad debris disks, which will be exciting to confirm and explore with future high-resolution observations.  

The cause of the gapped structure in debris disks is not well understood.  The main categories into which proposed mechanisms fall include: (1) gas-dust dynamics, (2) inheritance from the protoplanetary disk phase, and (3) dynamical interactions with a planet.  

\citet{lyra2013} note that a robust clumping instability can organize dust into multiple eccentric rings even in the absence of a planet.  The main prerequisite for the effect to operate is a gas-to-dust mass ratio $\gtrsim 1$.  Given our upper limit on the CO flux density of 40\,mJy\,km\,s$^{-1}$, the approximate upper limit on the CO mass, assuming LTE and an excitation temperature of 30\,K, is roughly $8.4\times 10^{-5}$\,$M_\earth$, using molecular data for CO v=0 drawn from the Cologne Database for Molecular Spectroscopy \citep{endres2016}.  With the best-fit dust mass from Table~\ref{table:MCMC} of 0.021\,M$_\earth$, the upper limit on the gas-to-dust mass ratio is $1.0\times 10^{-4}$ if we assume that the gas is dominated by CO, or 0.060 if we assume a protoplanetary-like composition with a CO/H$_2$ abundance ratio of $10^{-4}$, corresponding to a mass ratio of $\sim$0.0014.  Given these stringent limits, it is unlikely that gas dynamics are responsible for the double-ringed structure in this system.  

One way in which gas dynamics could conceivably play a role despite the lack of gas in the present-day debris disk is if the gapped structure is inherited from the protoplanetary disk phase.  The Disk Structures at High Angular Resolution Project (DSHARP) has characterized the locations of rings and gaps in a sample of 20 (18 of which are single) bright, nearby protoplanetary disks \citep{huang2018}.  The outer dust radius of HD~206893 would place it among the larger disks in the sample (3 of 20 DSHARP disks are as large or larger), and the {  central} gap radius of 78\,au would place it roughly in the 79th percentile of the 52 protoplanetary disk gaps identified by DSHARP.  The  structure of the HD~206893 system is therefore generally consistent with the distribution of disk radii and gap locations identified by DSHARP, {  though perhaps on the larger side especially given its spectral type.  It is also worth noting that the DSHARP sample is biased towards systems with high dust luminosities, and is almost certainly not representative}.  Of course, the origin of the gaps and rings in protoplanetary disks is also not well understood, so if the structure is inherited then at most we have gained some evidence that the dust rings in DSHARP do correspond to planetesimal rings.  Proposed mechanisms for generating the gaps in protoplanetary disks include chemical effects like snow lines \citep[e.g.,][]{banzatti2015,okuzumi2016}, magnetohydrodynamic instabilities of various flavors along with resulting gas pressure gradients \citep[e.g.,][]{johansen2009,bai2014,simon2014,flock2015,lyra2015}, and, of course, one or more (proto)planets \citep[e.g.][]{papaloizou1984,bryden1999,nelson2000}. At this point, there is enough evidence pointing to the presence of planets in the disk at young ages to make it likely that at least some of the gaps and rings in protoplanetary disks are indeed caused by planets \citep[e.g.,][]{isella2018,isella2019,pinte2018,teague2018}.  

The extent to which disk structure is likely to be retained between the protoplanetary and debris disk phase is still largely an open question.  Most theoretical studies of debris disk structure assume that the protoplanetary disk dust profile is not retained, and make predictions for debris disk structure that depend only on the locations of colliding planetesimals that produce the dust, along with the masses and orbital properties of the larger bodies that dynamically sculpt it (plus, for smaller grains, radiation pressure, stellar winds, and sometimes Poynting-Robertson drag, depending on the collision rate).  

In the absence of a reservoir of gas whose mass is comparable to that of the dust, it is difficult to break the radial symmetry of the disk without planets.  Even within the category of planet-sculpted gaps, there are multiple theoretical considerations that point to different masses and orbital properties of the underlying planetary system.  However, the most straightforward explanation is that a single additional unseen planet-mass companion, located within the gap, is responsible for clearing the dust at the location of the gap.  In that case, the semimajor axis and width of the gap encode the location and mass of the planet.  Ignoring the presence of the brown dwarf at 10\,au and using the best-fit values for {  $R_\mathrm{gap}$ and $\Delta R_\mathrm{gap}$ of 67\,au and 32\,au,} respectively, the three-body chaotic zone theory would predict that the gap is carved by a planet with semimajor axis {  $a_p = 80$\,au and mass ratio $\mu = 1.0\times 10^{-3}$, corresponding to a mass of 1.4\,M$_\mathrm{Jup}$,} assuming an orbital eccentricity of zero.  To derive these estimates, we use the general functional form from  \citet{morrison2015} where $\Delta a = C \mu^\beta a_p$, where $C$ and $\beta$ are constants tabulated separately for the inner and outer chaotic zone width for a range of ratios of the radius of the planet $R_p$ to the extent of the Hill sphere $R_H$; we use the values tabulated for $R_p/R_H = 0.001$. If the planet has significant eccentricity, then the width of the gap would be expected to increase as $1.8 e^{1/5} \mu^{1/5}$, at least above a critical eccentricity of {  $0.21\mu^{3/7} = 0.004$,} implying that a lower-mass planet on an eccentric orbit could be responsible for carving the observed {  31\,au} gap width \citep{mustill2012}.  Propagating uncertainties through to the mass estimate is non-trivial, not only due to the relative uncertainties on our measured values of the gap radius and depth, but also due to uncertainties on the parameters $C$ and $\beta$, the asymmetric nature of the chaotic zone, and the degeneracy between model parmeters like $p$ and the gap properties, so the value of {  1.4\,M$_\mathrm{Jup}$} should be considered a ballpark estimate subject to many systematic and relative uncertainties. 

More exotic possibilities have been invoked to explain the gapped structure in the HD~107146 disk, notably the possibility that both the inner disk edge and the gap could be carved by a single planet with mass comparable to the debris disk and eccentricity between 0.4-0.5, located interior to the inner edge of the broad debris belt \citep{pearce2015}.  That configuration was ultimately ruled out for the HD~107146 system by \citet{marino2018}.  It is similarly unlikely to apply to the HD~206893 system, because the location of the inner disk edge appears to be adjacent to the chaotic zone of the directly imaged brown dwarf, and the brown dwarf's mass is so much larger than that of the disk that the apsidal anti-alignment described for low-mass companions in \citet{pearce2015} would not occur.  Another possibility, of course, is that multiple planets could be carving the gap, in which case the mass of each planet would be substantially smaller than the {  1.4\,M$_\mathrm{Jup}$} that we derive for the single-planet case.  

\subsection{N-body Simulations of the Star, Brown Dwarf, and Putative Planet}

The chaotic zone theory on which we base our estimate of planet mass is fundamentally a 3-body result, but adding the dynamical influence of the brown dwarf (effectively treating the central star as a low-$\mu$ binary system) makes the system fundamentally a 4-body problem. We therefore conducted N-body simulations with and without the brown dwarf (the latter to verify that the gap depth and width were not significantly affected by the presence of the brown dwarf) to investigate whether the orbital properties of the brown dwarf have a detectable impact on the width of the gap.

In order to simulate gravitational interaction between the star, brown dwarf, putative planet, and dust particles in the disk, we used the N-body software \texttt{REBOUND} \citep{rein2012} with hybrid integrator \texttt{MERCURIUS} \citep{rein2019}, which switches from a fixed to variable timestep when any particle comes within a certain distance of a massive particle, here chosen to be three Hill radii. We chose the fixed timestep to be 4\% of the brown dwarf’s initial period, which is less than 8\% of every dust particle’s period when the simulation begins. Dust particles are treated in the test particle limit.
  
Particles are initially randomly distributed with a uniform distribution of semimajor axes between 8 and 158\,au, a uniform distribution of eccentricity {  from} 0 {  to} 0.02, and a uniform distribution of inclination {  from} 0\textsuperscript{\(\circ\)} {  to} 10\textsuperscript{\(\circ\)}. We placed 10$^4$ disk particles, adequate to sample the 150\,au span of the disk and recover smooth images of the final disk density distribution. We assumed a stellar mass of 1.32 M\textsubscript{\(\sun\)} and a stellar radius of 1.26 R\textsubscript{\(\sun\)} \citep{delorme2017}.  The brown dwarf is placed on a orbit with semimajor axis 10.4\,au and assumed to have a radius of 1.3\,R\textsubscript{Jup}. We assumed a planet mass of {  1.4\,M$_\mathrm{Jup}$} and a bulk density of {  1.3\,g\,cm$^{-3}$}. We placed the planet on a circular orbit at {  80\,au}. We integrate the system up to 10\,Myr, {  which is sufficiently long to capture many orbital timescales for all the bodies.} We then assign a weight to each particle based on its initial semimajor axis to mock up a surface density {  that follows the best-fit double power-law model parameters}. We ran a set of simulations, varying the brown dwarf mass between 12 and 50\,M\textsubscript{Jup} \citep[the range of plausible values from][]{grandjean2019} with a {  1.3\,M$_\mathrm{Jup}$} planet. We also investigated the effects of placing the brown dwarf on orbits with eccentricities of 0.03, 0.1, and 0.3, as well as introducing an orbital inclination to the plane of the planetesimal disk of 5\textsuperscript{\(\circ\)} or 10\textsuperscript{\(\circ\)}.  
 
\begin{figure}[ht!]
\centering
\gridline{\fig{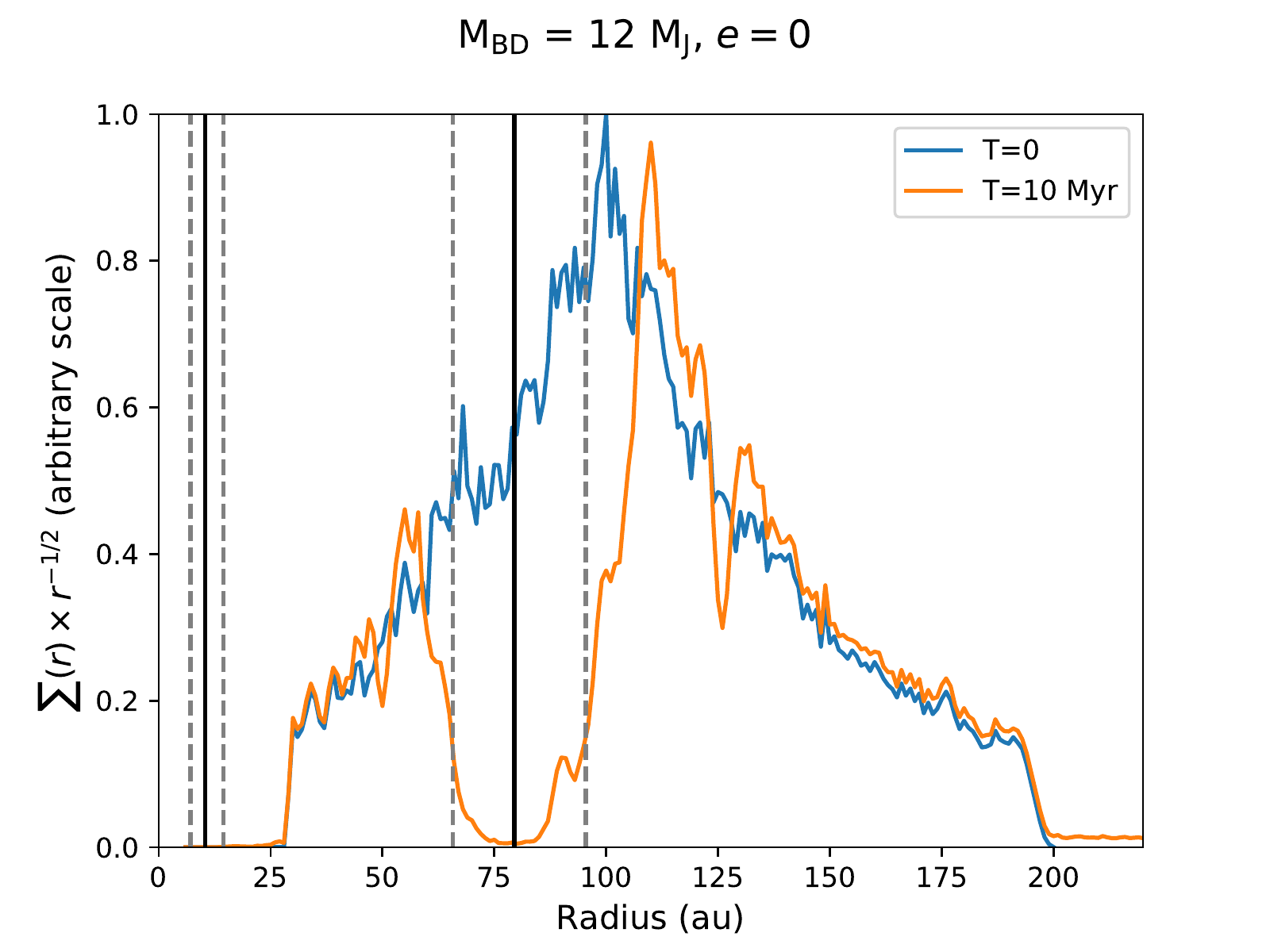}{0.48\textwidth}{}
          \fig{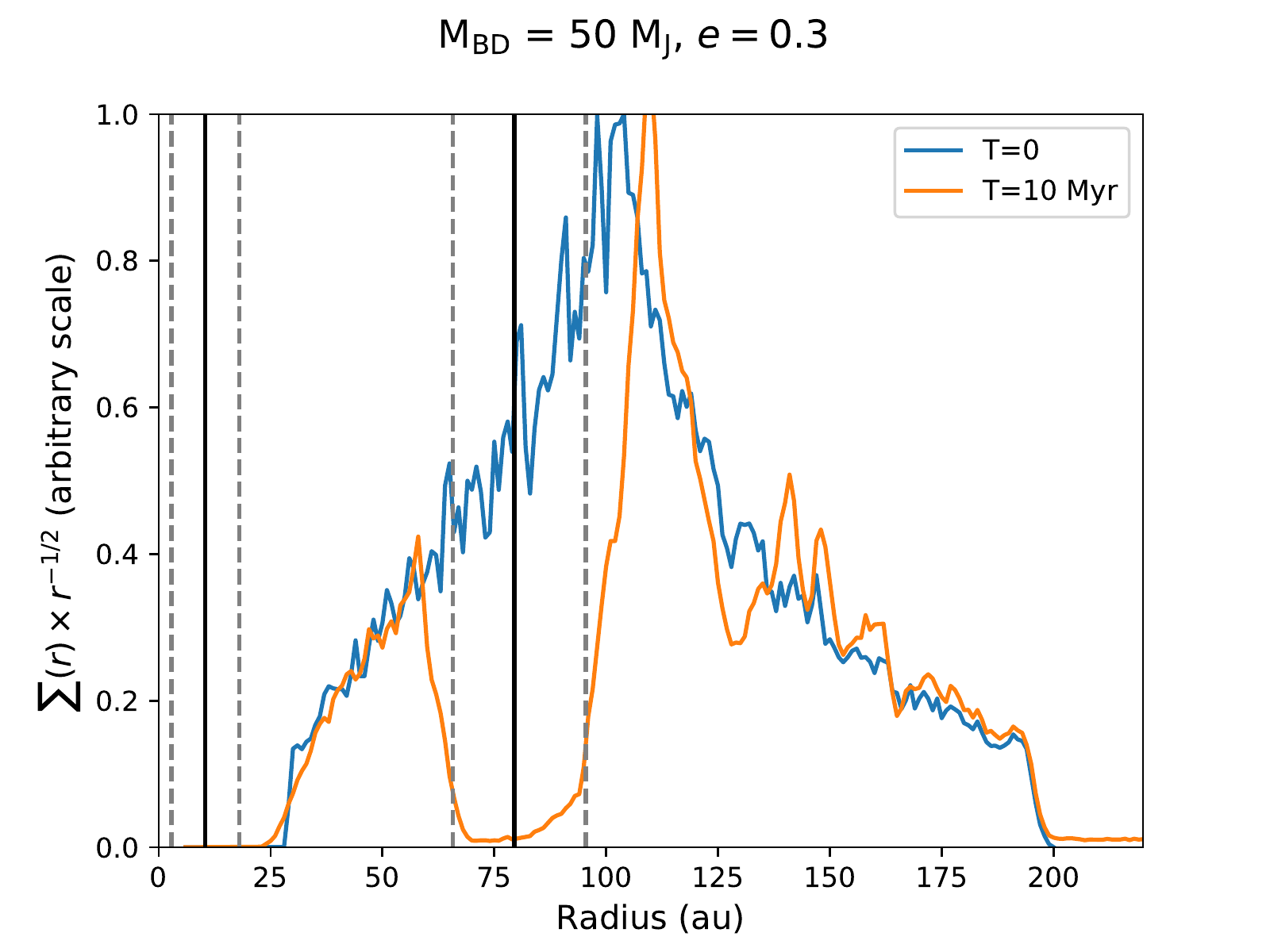}{0.48\textwidth}{}
          }
\caption{Surface brightness of particles in an N-body simulation of the HD 206893 system as a function of radius relative to the barycenter of the system.  These plots show the evolution of disk particles over 10\,Myr, with a planet on a circular orbit at {  80\,au}. The surface brightness assumes an initial surface density proportional to {  that derived for the double power-law model in Table~\ref{table:MCMC}} and a dust temperature profile proportional to $R^{-1/2}$. The dashed vertical lines represent the chaotic zone of the brown dwarf and the {  1.4\,M$_\mathrm{Jup}$} planet, calculated according to formulae from \citet{mustill2012,morrison2015}. The left panel shows a system with a 12 M\textsubscript{Jup} brown dwarf with 0 eccentricity. The right panel shows a system with a 50M\textsubscript{Jup} brown dwarf with 0.3 eccentricity.}
\label{fig:n-body}
\end{figure}

A summary of the properties of the brown dwarf and the measured gap width and depth based on the final particle distribution is presented in Table~\ref{tab:n-body}.  The gap width and depth were estimated by fitting a power law to the initial surface brightness distribution, subtracting the final profile from the fit to the initial profile, excluding the ``Trojan" particles at the center of the gap, and fitting a top hat function to the result.  {  Trojans were defined as any particle with a semimajor axis that falls within 1 Hill Radius of the semimajor axis of the planet.}  Two sample radial surface brightness profiles are presented in Figure~\ref{fig:n-body}.  The left panel shows the result for a low-mass (12\,M$_\mathrm{Jup}$) brown dwarf with zero eccentricity, and the right panel shows the result for a high-mass (50\,M$_\mathrm{Jup}$) brown dwarf with an eccentricity of 0.3.  In both panels, the solid vertical lines represent the location of the brown dwarf (near 10.4\,au) and planet (near 79\,au), and the adjacent dashed lines represent the extent of each body's chaotic zone, as calculated from the formulae presented in \citet{morrison2015} for the zero-eccentricity case and \citet{mustill2012} for the high-eccentricity case.  While the top-hat function fit to the surface brightness profile yields a slightly broader width of {  34\,au} than expected from the chaotic zone extent of {  32\,au}, the figures demonstrate that the gap edge is not well defined and the chaotic zone approximation is a good estimate for where the surface brightness starts to increase away from the influence of the companion. 

We therefore find that the N-body simulations confirm that the chaotic zone approximation is appropriate for most of the parameter space covered by current estimates of the properties of the brown dwarf companion. Even a high mass brown dwarf (50\,M\textsubscript{Jup}) with substantial eccentricity does not significantly perturb the planet’s orbit (and therefore the width and location of the gap estimated by chaotic zone theory). This result is consistent with a back-of-the-envelope estimate which places the secular timescale at $<10$\,Myr and the maximum eccentricity of the outer planet of 0.06 for a brown dwarf-to-planet mass ratio $>>1$.  The depth of the gap for the case of 15\,M$_\mathrm{Jup}$ is roughly {  94\%}, which is slightly shallower than the assumed 100\% depth of the gap in our ray-tracing code. Assuming a shallower gap would likely lead to a broader estimated width of the gap in the MCMC code, indicating that our estimate of the planet mass may be slightly too low.  However, since the N-body simulation includes Trojans that may or may not be realistic (for example, the model does not include collisional evolution or the formation of the planet's core, either of which might deplete particles within the gap).  

{  Finally, we then fixed the mass of the brown dwarf at 15 M$_\mathrm{Jup}$, on a circular orbit at 10.4\,au, and ran a set of simulations varying the planet mass between 0.5 and 50\,M$_\earth$.  We also investigated the effects of placing the planet with eccentricities of 0.03, 0.1, and 0.3, as well as introducing an orbital inclination to the plane of the disk of 5$^\circ$, 10$^\circ$, and 15$^\circ$.  A summary of the properties of the planet and the measured gap width and depth based on the final particle distribution is presented in Table~\ref{table:planet_m_e}.  The gap width calculated by the top-hat function fig continues to yield results higher than the chaotic zone estimate for all planet masses and eccentricities.  However, the changes in gap width generally conform to chaotic zone theory, as gap width increases with planet mass to the 0.27 power, close to the formula from \citep{morrison2015}.  The gap width increases proportionally to eccentricity to the 0.21 power, close to the formula from \citep{mustill2012}.  As expected, as mass increased the depth of the gap increased as well as the width.  As eccentricity increased, the depth of the gap decreased.  Inclining the planet's orbit had little effect on the width of the gap, although the depth of the gap slightly decreased.}

\begin{table*}[ht]
\centering
\caption{Gap Width and Depth with Varying Brown Dwarf Parameters}
\label{tab:n-body}
\begin{center}
\resizebox{\textwidth}{!}{
\begin{tabular}{lcccccccc}
\hline
BD Mass (M\textsubscript{Jup}) & BD Eccentricity & BD Inclination(\textsuperscript{\(\circ\)}) & Gap Width (au) & Gap Depth (\%) \\
\hline
12 & 0 & 0 & 34 & 95\\
15 & 0 & 0 & 35 & 94\\
30 & 0 & 0 & 35 & 97\\
50 & 0 & 0 & 34 & 96\\
50 & 0.03 & 0 & 34 & 98\\
50 & 0.1 & 0 & 34 & 96\\
50 & 0.3 & 0 & 35 & 99 \\
50 & 0 & 5 & 34 & 95 \\
50 & 0 & 10 & 34 & 97 \\
No BD & & & 34 & 98 \\
\hline
\end{tabular}
}
\end{center}
\end{table*}

\begin{table*}[ht]
\centering
\caption{Gap Width and Depth with Varying Planet Parameters}
\label{table:planet_m_e}
\begin{center}
\resizebox{\textwidth}{!}{
\begin{tabular}{lcccccccc}
\hline
Planet   Mass (M\textsubscript{\(\Earth\)}) & Planet Eccentricity & Planet Inclination (\textsuperscript{\(\circ\)}) & Gap Width (AU) & Gap Depth (\%) \\
\hline
0.5 & 0 & 0 & \multicolumn{2}{l}{Gap Undetectable} \\
1 & 0 & 0 & 7 & 40 \\
5 & 0 & 0 & 11 & 51 \\
10 & 0 & 0 & 13 & 58 \\
15 & 0 & 0 & 15 & 62 \\
20 & 0 & 0 & 16 & 60 \\
25 & 0 & 0 & 16 & 59 \\
30 & 0 & 0 & 18 & 64 \\
35 & 0 & 0 & 18 & 66 \\
40 & 0 & 0 & 19 & 71 \\
45 & 0 & 0 & 20 & 71 \\
50 & 0 & 0 & 20 & 72 \\
100 & 0 & 0 & 24& 86\\
200 & 0 & 0 & 29& 93\\
400 & 0 & 0 & 33& 97\\
445 & 0 & 0 & 34 & 96 \\
445 & 0.01 & 0 & 34 & 95 \\
445 & 0.03 & 0 & 35 & 95 \\
445 & 0.1 & 0 & 44 & 89 \\
445 & 0.3 & 0 & Gap Undetectable\\
445 & 0 & 5 & 34 & 96 \\
445 & 0 & 10 & 35 & 94 \\
445 & 0 & 15 & 34 & 92 \\
\hline
\end{tabular}
}
\end{center}
\end{table*}

\section{Summary and Conclusions}
\label{sec:conclusions}

As one of only two known systems to host a brown dwarf orbiting within a debris ring, HD~206893 presents a rare and valuable opportunity to study companion-disk  interactions and place dynamical constraints on the mass of a directly imaged companion.  The ALMA observations at a wavelength of 1.3\,mm presented here spatially resolve the radial structure of the disk, revealing a broad distribution of planetesimals extending from radii of $<51$\,au to  $194^{+13}_{-2}$\,au, with statistically significant (according to the AIC/BIC) evidence for a gap in the dust disk with inner radius  $63^{+8}_{-16}$\,au and width  $31^{+11}_{-7}$\,au.  

The inner radius of the disk is not resolved by the current ALMA observation of the system, allowing us to place only a modest upper limit on the mass of the companion of $<1170$\,M$_\mathrm{Jup}$.  The serendipitous discovery of a gap in the disk marks the fourth time that a radial gap has been discovered within a broad debris belt at millimeter wavelengths, among the $\sim$6 systems that have so far been surveyed at sufficient resolution and sensitivity to detect such a gap.  While the origin of gapped structure is still unclear, in this case the low limit on the gas mass in the system renders pressure gradients due to residual disk gas an unlikely explanation, so the gap must either have been inherited from the protoplanetary disk phase or carved by one or more additional, unseen companions at larger separation in the system.  If the gap is carved by a single planet on a circular orbit, chaotic zone theory predicts that it should have a mass of  1.4\,M$_\mathrm{Jup}$ at a semimajor axis of 79\,au.  

Future ALMA observations at higher angular resolution have the potential to not only place valuable dynamical constraints on the mass of the brown dwarf companion by measuring the location of the inner disk edge, but can also better constrain the properties of the putative planet by measuring the gap width and depth.  Surveying the radial structure of broad debris disks at millimeter wavelengths has the potential to distinguish between scenarios in which the gapped structure is inherited from the protoplanetary disk and scenarios in which it is actively carved by unseen planets, while also providing guidance and insight for future direct imaging surveys for planets in debris-rich systems. 

\acknowledgments
We express appreciation to Rebekah Dawson for advice about the N-body simulations.  A.N. is sponsored by Wesleyan University’s Research in the Sciences Fellowship and Wesleyan University's Quantitative Analysis Center Apprenticeship. A.M.H. is supported by a Cottrell Scholar Award from the Research Corporation for Science Advancement. A.M. acknowledges the support of the Hungarian National Research, Development  and  Innovation  Office  NKFIH  Grant  KH-130526.  K.Y.L.S. acknowledges the partial support from NASA ADAP grant NNX15AI86G.  This paper makes use of the following ALMA data: ADS/JAO.ALMA\#2018.1.00193.S. ALMA is a partnership of ESO (representing its member states), NSF (USA) and NINS (Japan), together with NRC (Canada), MOST and ASIAA (Taiwan), and KASI (Republic of Korea), in cooperation with the Republic of Chile. The Joint ALMA Observatory is operated by ESO, AUI/NRAO and NAOJ. 

This work has made use of data from the European Space Agency (ESA) mission
{\it Gaia} (\url{https://www.cosmos.esa.int/gaia}), processed by the {\it Gaia}
Data Processing and Analysis Consortium (DPAC,
\url{https://www.cosmos.esa.int/web/gaia/dpac/consortium}). Funding for the DPAC
has been provided by national institutions, in particular the institutions
participating in the {\it Gaia} Multilateral Agreement.

\textit{Software:} \texttt{Astropy} \citep{astropy2013}, \texttt{CASA} \citep{mcmullin2007}, \texttt{emcee} \citep{foreman-mackey2013}, \texttt{Matplotlib} \citep{hunter2007}, \texttt{MIRIAD} \citep{sault1995}, \texttt{NumPy} \citep{vanderWalt2011},  \texttt{Pandas} \citep{mckinney2010}, \texttt{Uncertainties}, \url{http://pythonhosted.org/uncertainties}

\appendix
\label{appendix:MCMC}

\begin{figure}[ht!]
\centering
\includegraphics[angle=0,width=\textwidth]{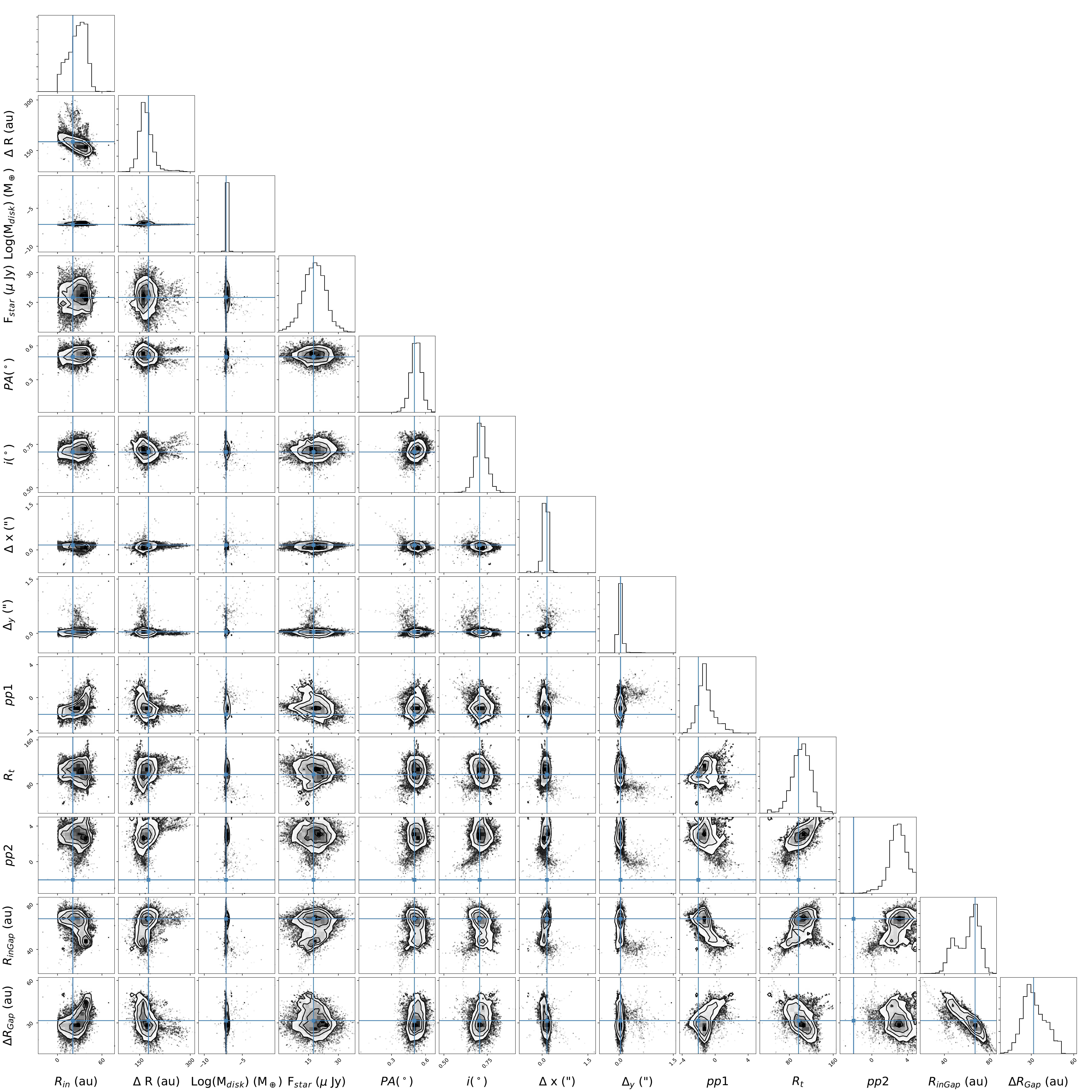}
\caption{Corner plot for the MCMC chain for the double power-law model with a gap, after removing burn-in.  Histograms for each parameter are plotted at the top of the corresponding column, while the plots for the rest of the grid show the distribution of walkers across slices through parameter space for the corresponding pair of parameters.  The best-fit value for each parameter is plotted with a blue line. }
\label{figure:corner_doublepp}
\end{figure}

\begin{figure}[ht!]
\centering
\includegraphics[angle=0,width=\textwidth]{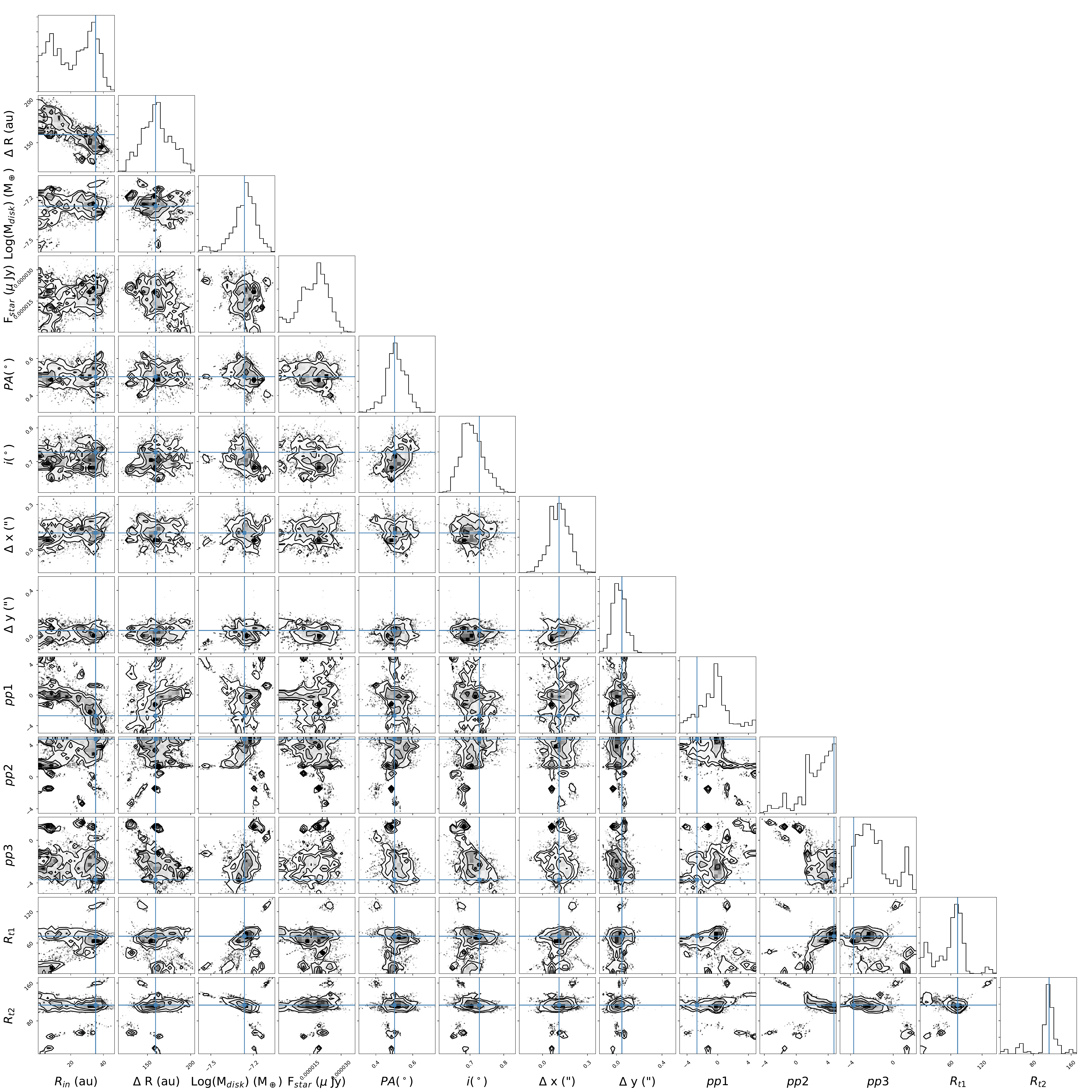}
\caption{Corner plot for the MCMC chain for the triple power-law model after removing burn-in.  Symbols as in Figure~\ref{figure:corner_doublepp}. }
\label{figure:corner_triplepp}
\end{figure}

In this Appendix, we show the corner plots for the MCMC chains for the double power law model with a gap (Figure~\ref{figure:corner_doublepp}) and for the triple power law model (Figure~\ref{figure:corner_triplepp}).  The corner plots show a histogram for each parameter at the top of each column, and slices through parameter space for the rest of the grid.  The best-fit value of each parameter is marked with a blue line.  

\bibliographystyle{apj}{}
\bibliography{hd206893}

\end{document}